\documentclass[12pt]{article}

\usepackage{amsmath}
\usepackage{amssymb}
\usepackage{amsthm}
\usepackage{array}
\usepackage[footnotesize,bf]{caption}
\usepackage{color}
\usepackage{enumitem}
\usepackage{eurosym}
\usepackage[top=2.4cm,left=2.7cm,right=2.7cm,bottom=3cm]{geometry}
\usepackage{graphicx}
\usepackage{lineno}
\usepackage[sc]{mathpazo}
\usepackage{mathtools}
\usepackage{multirow}
\usepackage{setspace}
\usepackage{times}
\usepackage{titlesec}
\usepackage[normalem]{ulem}
\usepackage{verbatim}
\usepackage{xr}

\smallskip

\definecolor{mygreen}{rgb}{0.0, 0.6, 0.0}

\titleformat{\section}{\sffamily \fontsize{13}{16}\bfseries}{\thesection}{1em}{}
\titleformat{\subsection}{\sffamily \fontsize{11.5}{11.5}\bfseries}{\thesubsection}{1em}{}

\newcommand*\patchAmsMathEnvironmentForLineno[1]{%
	\expandafter\let\csname old#1\expandafter\endcsname\csname #1\endcsname
	\expandafter\let\csname oldend#1\expandafter\endcsname\csname end#1\endcsname
	\renewenvironment{#1}%
	{\linenomath\csname old#1\endcsname}%
	{\csname oldend#1\endcsname\endlinenomath}}%
\newcommand*\patchBothAmsMathEnvironmentsForLineno[1]{%
	\patchAmsMathEnvironmentForLineno{#1}%
	\patchAmsMathEnvironmentForLineno{#1*}}%
\AtBeginDocument{%
	\patchBothAmsMathEnvironmentsForLineno{equation}%
	\patchBothAmsMathEnvironmentsForLineno{align}%
	\patchBothAmsMathEnvironmentsForLineno{flalign}%
	\patchBothAmsMathEnvironmentsForLineno{alignat}%
	\patchBothAmsMathEnvironmentsForLineno{gather}%
	\patchBothAmsMathEnvironmentsForLineno{multline}%
}

\theoremstyle{definition}

\title{\begin{center} \bfseries \singlespacing
Learning enables adaptation in cooperation for multi-player stochastic games
\end{center}}
\author{\parbox[c]{16cm}{\onehalfspacing \normalsize \centering ~\\[-0.4cm] Feng Huang$^{1,2}$
\quad Ming Cao$^{2,\ast}$
\quad Long Wang$^{1}$\footnote{Corresponding authors: Long Wang (\texttt{longwang@pku.edu.cn}) and Ming Cao (\texttt{m.cao@rug.nl}).}\\ \quad\\ \footnotesize
$^{1}$Center for Systems and Control, College of Engineering, Peking University, Beijing 100871, P. R. China \\
$^{2}$Center for Data Science and System Complexity, Faculty
of Science and Engineering, University of Groningen, Groningen 9747 AG,
The Netherlands \\[0.2cm]}
\date{}
}

\begin{document}

\maketitle

\begin{abstract}
Interactions among individuals in natural populations often occur in a dynamically changing environment. Understanding the role of environmental variation in population dynamics has long been a central topic in theoretical ecology and population biology. However, the key question of how individuals, in the middle of challenging social dilemmas (e.g., the ``tragedy of the commons''), modulate their behaviors to adapt to the fluctuation of the environment has not yet been addressed satisfactorily. Utilizing evolutionary game theory and stochastic games, we develop a game-theoretical framework that incorporates the adaptive mechanism of reinforcement learning to investigate whether cooperative behaviors can evolve in the ever-changing group interaction environment. When the action choices of players are just slightly influenced by past reinforcements, we construct an analytical condition to determine whether cooperation can be favored over defection. Intuitively, this condition reveals why and how the environment can mediate cooperative dilemmas. Under our model architecture, we also compare this learning mechanism with two non-learning decision rules, and we find that learning significantly improves the propensity for cooperation in weak social dilemmas, and, in sharp contrast, hinders cooperation in strong social dilemmas. Our results suggest that in complex social-ecological dilemmas, learning enables the adaptation of individuals to varying environments.
\end{abstract}

\singlespacing

\section{Introduction}
Throughout the natural world, cooperating through enduring a cost to endow unrelated others with a benefit is evident at almost all levels of biological organisms, from bacteria to primates~\cite{west2007evolutionary}. This phenomenon is especially true for modern human societies with various institutions and nation-states, in which cooperation is normally regarded as the first choice to cope with some major global challenges, such as curbing global warming~\cite{pachauri2010climate,milinski2008collective} and governing the commons~\cite{ostrom2015governing}. However, the mechanism underlying cooperative behavior has perplexed evolutionary biologists and social economists for a long time~\cite{colman2006puzzle,nowak2006five}. Since according to the evolutionary theory of ``survival of the fittest'' and the hypothesis of Homo economicus, this costly prosocial behavior will be definitively selected against and should have evolved to be dominated by selfish act~\cite{dawkins2016selfish}.
\par
To explain how cooperation can evolve and be maintained in human societies or other animal groups, a large body of theoretical and experimental models have been put forward based on evolutionary game theory~\cite{nowak2006five,smith1982evolution,hofbauer1998evolutionary} and social evolution theory~\cite{hamilton1964genetical}. Traditionally, the vast majority of the previous work addressing this cooperative conundrum concentrates on the intriguing paradigm of a two-player game with two strategies, the prisoner's dilemma~\cite{nowak2006five,szabo2007evolutionary}. Motivated by abundant biological and social scenarios where interactions frequently occur in a group of individuals, its multi-person version -- the public goods game -- has attracted much attention in recent years~\cite{archetti2012game}. Meanwhile, it also prompts a growing number of researchers to devote to studying multi-player games and multi-strategy games~\cite{gokhale2010evolutionary,tarnita2011multiple,wu2013dynamic,pena2016ordering,mcavoy2016structure}. However, these prominent studies implicitly assume, as most of the canonical work does, that the game environment is static and independent of players' actions. In other words, in these models, how players act by choosing game-play strategies only affects the strategic composition in the population, but the game environment itself is not influenced. As a result, a single fixed game is played repeatedly. Of course, this assumption is well grounded, if the timescale of interest (e.g., the time to fixation or extinction of a species) is significantly shorter than that of the environmental change. For most of realistic social and ecological systems, however, it seems to be too idealized. Hence, an explicit consideration of environmental change is needed. A prototypical instance is the overgrazing of common pasture lands~\cite{hardin1968tragedy}, where the depleted state may force individuals to cooperate and accordingly the common-pool resources will increase, whereas the replete state may induce defection and the common-pool resources will decrease~\cite{weitz2016oscillating,hilbe2018evolution}. Other examples also exist widely across scales from small-scale microbes to large-scale human societies~\cite{estrela2019environmentally}. A common feature of these examples is the existence of the feedback loop where individual behaviors alter environmental states, and are influenced in turn by the changed environment~\cite{weitz2016oscillating,tilman2020evolutionary}.
\par
Although the effect of environmental variations on population dynamics has long been recognized in theoretical ecology and population biology~\cite{macarthur1970species,levins1968evolution,ROSENBERG20201}, it is only recently that there has been a surge of interest in constructing game-environment feedbacks~\cite{weitz2016oscillating,hilbe2018evolution,tilman2020evolutionary,ashcroft2014fixation,chen2018punishment,Su19Evolutionary,hauert2019asymmetric}
to understand the puzzle of cooperation. Different from the conventional settings in evolutionary game theory~\cite{smith1982evolution,hofbauer1998evolutionary}, the key conceptual innovation of these work is the introduction of multiple games~\cite{hashimoto2006unpredictability,venkateswaran2019evolutionary}, evolving games~\cite{stewart2014collapse}, dynamical system games~\cite{akiyama2000dynamical}, or stochastic games~\cite{shapley1953stochastic,neyman2003stochastic}. By doing so, the players' payoff depends on not only strategic interactions but also the environmental state, and meanwhile, the fluctuation of the environment will be subject to the actions adopted by players. So, the consideration of a dynamic game environment for the evolution of cooperation has as least two significant implications. On the one hand, it vastly expands the existing research scope of evolutionary game theory by adding a third dimension (multiple games) to the previous two-dimension space (multiple players and multiple strategies)~\cite{venkateswaran2019evolutionary}. In other words, this extension generalizes the existing framework to encompass a broader range of scenarios. On the other hand, the new key component, environmental feedbacks~\cite{weitz2016oscillating,tilman2020evolutionary}, is integrated seamlessly into the previous theoretical architecture.
\par
While these promising studies primarily focused on pre-specified or pre-programmed behavioral policies to analyze the interdependent dynamics between individual behaviors and environmental variations, the key question of how individuals adjust their behaviors to adapt to the changing environment has not yet been sufficiently addressed. In fact, when confronting complex biotic and abiotic environmental fluctuations, how organisms adaptively modulate their behaviors is of great importance for their long-term survival efforts~\cite{levins1968evolution,meyers2002fighting}. For example, those plants growing in the lower strata of established canopies can adjust their stem elongation and morphology in response to the spectral distribution of radiation, especially the ratio of red to far-red wavelength bands~\cite{ballare1990far}; in arid regions, bee larvae, as well as angiosperm seeds, strictly comply with a bet-hedging emergence and germination rule such that reproduction activities are only limited to a short period of time following the desert rainy season~\cite{danforth1999emergence}. Particularly, as an individual-level adaptation, learning through reinforcement is a fundamental cognitive or psychological mechanism used by humans and animals to guide action selections in response to the contingencies provided by the environment~\cite{thorndike1970animal,dayan2008reinforcement,sutton2018reinforcement}. Employing the experience gained from historical interactions, individuals always tend to reinforce those actions that will increase the probability of rewarding events and lower the probability of aversive events. Although this learning principle has become a central method in various disciplines, such as artificial intelligence~\cite{sutton2018reinforcement,bu2008comprehensive}, neurosicence~\cite{dayan2008reinforcement}, learning in games~\cite{fudenberg1998theory}, and behavioral game theory~\cite{camerer2011behavioral}, there is still a lack of the theoretical understanding of how it guides individuals to make decisions in order to resolve cooperative dilemmas.
\par
In the present work, we develop a general framework to investigate whether cooperative behaviors can evolve by learning through reinforcement in constantly changing multi-player game environments. To characterize the interplay between players' behaviors and environmental variations, we propose a normative model of multi-player stochastic games, in which the outcome of one's choice relies on not only the opponents' choices but also the current game environment. Moreover, we use a social network to capture the spatial interactions of individuals. Instead of using a pre-specified pattern, every decision-maker in our model learns to choose a behavioral policy by associating each game outcome with reinforcements. By doing so, our model not only considers the environmental feedback, but also incorporates a cognitive or psychological feedback loop (i.e., players' decisions determine their payoffs in the game, and in turn are affected by the payoffs). When selection intensity is so weak that the action choices of players are just slightly influenced by past reinforcements, we derive the analytical condition that allows for cooperation to evolve under the threat of the temptation to defection. Through extensive agent-based simulations, we validate the effectiveness of the closed-form criterion in well-mixed and structured populations. Also, we compare the learning mechanism with two non-learning decision rules, and interestingly, we find that learning markedly improves the propensity for cooperation in weak social dilemmas whereas hinders cooperation in strong social dilemmas. Furthermore, when the game is not stationary, we analyze how cooperation co-evolves with environmental states and the effect of external incentives on the cooperative evolution by agent-based simulations.

%
%

\section{Model and Methods}
\subsection{Model}
We consider a finite population of $N$ individuals living in an evolving physical or social environment. The population structure describing how individuals interact with their neighbors is characterized by a network, where nodes represent individuals and edges indicate interactions. When individuals interact with their neighbors, only two actions, cooperation ($C$) and defection ($D$), are available, and initially, every individual is initialized with a random action in the set $\mathcal{A}=\{C, D\}$ with a certain probability. In each time step, one individual is chosen randomly from the population to be the focal player, and then its $d-1$ neighbors as co-players are selected at random to form a $d$-player ($d\geq 3$) stochastic game~\cite{shapley1953stochastic,neyman2003stochastic}. To ensure that the game can always be organized successfully, we assume that each individual in the population has at least $d-1$ neighbors. Denote the possible number of $C$ players among $d-1$ co-players by the set $\mathcal{J}\triangleq\{0,1,\ldots,d-1\}$, and possible environmental states by the set $\mathcal{S}\triangleq\{s^1,s^2,\ldots,s^M\}$, where $s^i$, $i=1,2,\ldots,M$, represents the environmental state of type $i$. Then, depending on the co-players' configuration $j\in \mathcal{J}$ and the environmental state $s\in \mathcal{S}$ in the current round, each player will gain a payoff
given in Table~\ref{table1}. Players who take action $C$ will get a payoff $\mathrm{a}_{j}(s)\in\mathbb{R}$, whereas those who take action $D$ will get a payoff $\mathrm{b}_{j}(s)\in\mathbb{R}$, where $\mathbb{R}$ represents the set of real numbers. Players update their actions asynchronously; that is, in each time step, only the focal player updates its action, and other individuals still use the actions in the previous round. More specifically, to prescribe the updating rule, we define the policy $\pi(s,j,a;\theta,\beta)$ to quantify the probability that action $a$ is chosen by the focal player when there are $j$ opponents taking action $C$ among $d-1$ co-players in the environmental state $s\in\mathcal{S}$, where $\theta\in \mathbb{R}^L$ is the column parameter vector of $L$-dimension ($L\ll M$ to reduce dimensions) used for updating the policy by learning through reinforcement, and $\beta\in[0,+\infty)$ is the selection intensity~\cite{nowak2004emergence}, also termed the adaptation rate~\cite{sato2005stability}, which captures the effect of past reinforcements on the current action choice.
\begin{table}[!h]
\centering
\caption{Payoff table of the $d$-player stochastic game.}
\begin{tabular}{cccccc}
\hline
\hline
 Number of $C$ co-players & $d-1$ & \ldots & $j$ & \ldots & $0$ \\
\hline
$C$ & $\mathrm{a}_{d-1}(s)$  & \ldots & $\mathrm{a}_j(s)$ & \ldots & $\mathrm{a}_0(s)$ \\
$D$ & $\mathrm{b}_{d-1}(s)$  & \ldots & $\mathrm{b}_j(s)$ & \ldots & $\mathrm{b}_0(s)$ \\
\hline
\hline
\end{tabular}
\label{table1}
\end{table}
\par
After each round, players' decisions regarding whether to cooperate or defect in the game interaction will not only influence their immediate payoffs but also the environmental state in the next round. That is to say, the probability of the environmental state in the next round is conditioned on the action chosen by the focal player and the environmental state in the current round. Without loss of generality, we here assume that the dynamics of environmental states $\{s_t\}$ obey an irreducible and aperiodic Markov chain, which thus possesses a unique stationary distribution. Also, from Table~\ref{table1}, it is clear that the payoff of each player is a function of the environmental state. Therefore, when the environment transits from one state to another, the type of the normal-form (multi-player) game defined by the payoff table will be altered accordingly.
\par
\begin{figure}[!htb]
  \centering
  \includegraphics[width=\hsize]{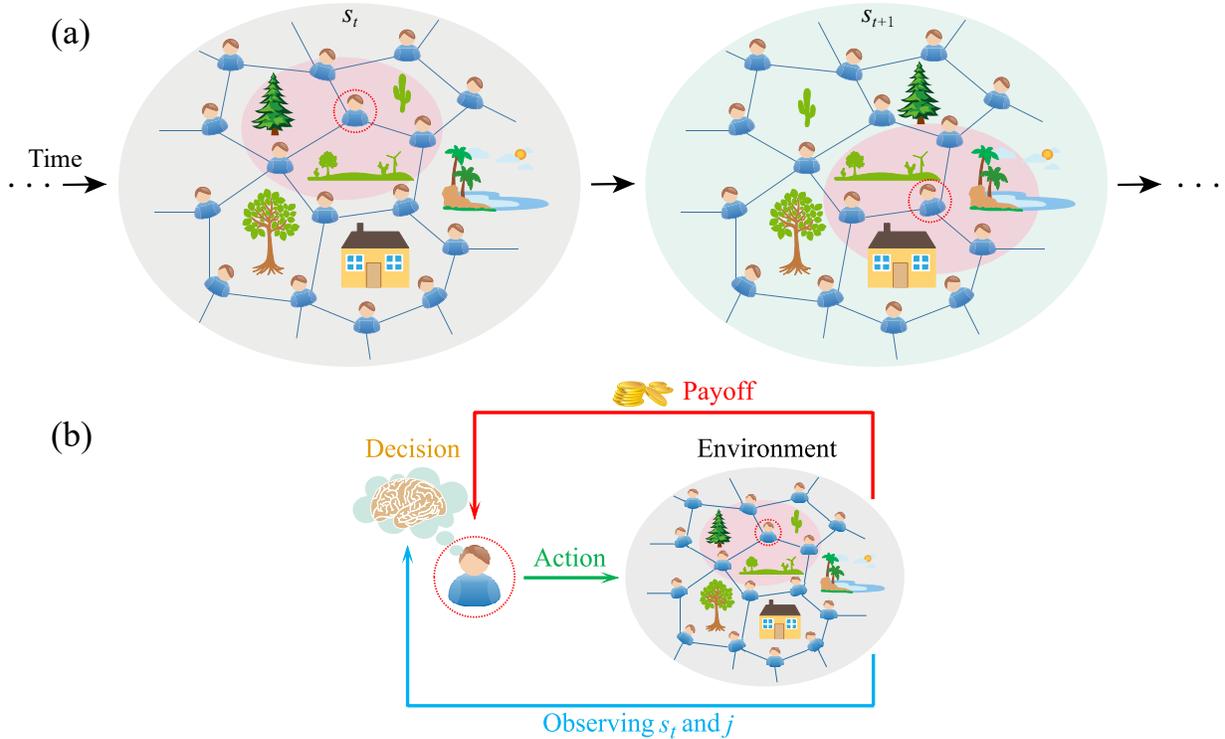}
  \caption{Illustration of evolutionary dynamics for $4$-player stochastic games in the structured population. (a), At a time step $t$, a random individual is chosen as the focal player (depicted by the dashed red circle), and then its $3$ neighbors are selected randomly as co-players to form a $4$-player game (because the focal player only has $3$ neighbors, all of them are chosen.), which is depicted by the light magenta shaded area. Conditioned on the focal player's action and the environmental state $s_t$ at time $t$, the environmental state at time $t+1$ will change to $s_{t+1}$ with a transition probability. Similarly, a new round of the game will be reorganized at time $t+1$. This process is repeated infinitely. (b), At time $t$, after perceiving the environmental state $s_t$ and the co-players' configuration $j$, the focal player uses the policy $\pi$ to determine which action to be taken, whereas its co-players still use their previous actions in the last round. At the end of this round, each player will gain a payoff. The received payoff will play the role of the feedback signal, and render the updating of the policy used by the focal player.
  }
  \label{fig1}
\end{figure}

The emergence of the new environmental state in the next round, apart from influencing the game type, may also trigger players to adjust their behavioral policies. This is because those previously used decision-making schemes may not be appropriate anymore in the changed environment. We here consider a canonical learning mechanism, actor-critic reinforcement learning~\cite{thorndike1970animal,dayan2008reinforcement,sutton2018reinforcement}, to characterize the individual adaptation to the fluctuating environment. Specifically, after each round, the players' payoffs received from the game interaction will play a role of the incentive signal of the interactive scenario. If one choice gives rise to a higher return in a certain scenario, then it will be reinforced with a higher probability in the future when encountering the same situation again. In contrast, those choices resulting in lower payoffs will be weakened gradually. Technically, this process is achieved via updating the learning parameter $\theta$ of the policy after each round (see Methods for more details). In the successive round, the acquired experience will be shared within the population and the updated policy will be reused by the newly chosen focal player to determine which action to be taken. In a similar way, this dynamical process of game formation and policy updating is repeated infinitely (Fig.~\ref{fig1}).

\subsection{Methods}
\subsubsection{Actor-critic reinforcement learning}
As the name suggests, the architecture of the actor-critic reinforcement learning consists of two modules. The actor module maintains and learns the action policy. Generally, there are two commonly used forms, $\epsilon$-greedy and Boltzmann exploration~\cite{sutton2018reinforcement,bu2008comprehensive}. Here, we adopt the latter for convenience, and consider the following Boltzmann distribution with a linear combination of features,
\begin{equation}\label{eq1}
  \pi(s,j,a;\theta,\beta)=\frac{e^{\beta\theta^T\phi_{s,j,a}}}
  {\sum_{b\in\mathcal{A}}e^{\beta\theta^T\phi_{s,j,b}}},\ \forall s\in \mathcal{S}, j\in\mathcal{J}, a\in \mathcal{A},
\end{equation}
where $\phi_{s,j,a}\in \mathbb{R}^L$ is the column feature vector with the same dimension of $\theta$, which is handcrafted to capture the important features when a focal player takes action $a$ given the environmental state $s$ and the number of $C$ players $j$ among its $d-1$ co-players. For the construction of the feature vector, there are many options, such as polynomials, Fourier basis, radial basis functions, and artificial neural networks~\cite{sutton2018reinforcement}. As mentioned in the Model, $\beta$ controls the selection intensity, or equivalently the adaptation rate. If $\beta\rightarrow 0$, it defines a weak selection and the action choice is only slightly affected by past reinforcements. When $\beta=0$, in particular, players choose actions with uniform probability. In contrast, if $\beta\rightarrow +\infty$, the action with the maximum $\theta^T\phi_{s,j,a}$ will be exclusively selected.
\par
Another module is the critic, which is used for learning an appropriate evaluation of the policy. For the long-run expected return of the policy per step, we evaluate it by defining a function $\rho(\pi)$,
\begin{equation}
  \rho(\pi)\triangleq\lim_{t\rightarrow \infty}\frac{1}{t}\mathbb{E}\{r_1+r_2+\cdots+r_t|\pi\},
\end{equation}
where $r_{t+1}\in\{\mathrm{a}_{d-1}(s),\ldots,\mathrm{a}_{0}(s),\mathrm{b}_{d-1}(s),\ldots,\mathrm{b}_{0}(s)\}$ is a random variable which denotes the payoff of the focal player at time $t\in\{0,1,2,\ldots\}$. In particular, if one denotes the probability when starting from the initial state $s_0$ the environmental state at time $t$ is $s_t$ under the policy $\pi$ by $Pr\{s_t=s|s_0,\pi\}$, and the average probability that all possible individuals chosen as the focal player encounter $j$ opponents taking action $C$ among $d-1$ co-players by $p_{\cdot j}$, then $\rho(\pi)$ can be given by
\begin{equation}\label{eq3}
  \rho(\pi)=\sum_{s\in\mathcal{S}} d^\pi(s)\sum_{j\in\mathcal{J}}p_{\cdot j}\sum_{a\in\mathcal{A}}\pi(s,j,a;\theta,\beta)\mathcal{R}_{s,j}^a,
\end{equation}
where $d^\pi(s)=\lim_{t\rightarrow \infty}Pr\{s_t=s|s_0,\pi\}$ is the stationary distribution of environmental states under the policy $\pi$; $\mathcal{R}_{s,j}^a$ is the payoff of the focal player when it takes action $a$ given the environmental state $s$ and the number of $C$ players $j$ among $d-1$ co-players, which can be given by
\begin{equation}\label{eq4}
\begin{split}
\mathcal{R}_{s,j}^a=
  \left\{
  \begin{array}{ll}
    \mathrm{a}_j(s), & \hbox{if }a=C; \\
    \mathrm{b}_j(s), & \hbox{if } a=D.
  \end{array}
\right.
\end{split}
\end{equation}
\par
Moreover, to evaluate the long-term accumulative performance of the policy, we define a Q-value function,
\begin{equation}\label{eq5}
  Q^\pi(s,j,a)\triangleq\sum_{t=1}^{\infty}\mathbb{E}\{r_t-\rho(\pi)
|s_0=s,j_0=j,a_0=a,\pi\},\ \forall s\in\mathcal{S}, j\in\mathcal{J}, a\in\mathcal{A},
\end{equation}
which is a conditional value dependent on the initial action $a_0=a$, environmental state $s_0=s$, and the number of $C$ players $j_0=j$ among $d-1$ co-players at time $t=0$. Since the space of the environmental state is usually combinatorial and thus extremely large in many game scenarios, it is not possible to calculate the Q-value function exactly for every environmental state, even in the limit given infinite time and data~\cite{sutton2018reinforcement}. Typically, one effective way to deal with this problem is to find a good approximation of the value function using limited computational resources. To this end, we approximate the Q-value function by a linear estimator~\cite{sutton2000policy,konda2000actor}, $f_w(s,j,a)$, given by
\begin{equation}\label{eq6}
\begin{split}
  f_w(s,j,a)&=w^T[\frac{\partial \pi(s,j,a;\theta,\beta)}{\partial \theta}\frac{1}{\pi(s,j,a;\theta,\beta)}] \\
  &=w^T[\phi_{s,j,a}
-\sum_{b\in\mathcal{A}}\pi(s,j,b;\theta,\beta)\phi_{s,j,b}]\beta,
\end{split}
\end{equation}
where $w\in \mathbb{R}^L$ is the column parameter vector used for updating the estimator.
\par
After a round $t$, depending on the payoff $r_{t+1}$ received by the focal player in the game, the policy of the focal player and the estimator of the Q-value function will be updated simultaneously via the following algorithm (see Supporting Information SI.1 for the algorithm derivation),
\begin{equation}\label{eq7}
\begin{split}
  w_{t+1}&=w_t+\alpha_t[r_{t+1}-\bar{R}_t+f_{w_t}(s_{t+1},j_{t+1},a_{t+1})
  -f_{w_t}(s_{t},j_{t},a_{t})]\frac{\partial f_{w_t}(s_t,j_t,a_t)}{\partial w_t}, \\
  \theta_{t+1}&=\theta_{t}+\gamma_t\frac{\partial \pi(s_t,j_t,a_t;\theta_t,\beta)}{\partial \theta_t}\frac{1}{\pi(s_t,j_t,a_t;\theta_t,\beta)}f_{w_t}(s_t,j_t,a_t),
\end{split}
\end{equation}
where $\bar{R}_t$ is the estimation of $\rho(\pi)$, and iterates through $\bar{R}_{t+1}=\bar{R}_t+[r_{t+1}-\bar{R}_t]/(t+1)$ and $\bar{R}_0=0$, $t=0,1,2,\ldots$; $\alpha_t$ and $\gamma_t$ are learning step-sizes which are positive, non-increasing for $\forall t$, and satisfy $\sum_{t}\alpha_t=\sum_{t}\gamma_t=\infty$, $\sum_{t}\alpha_t^2<\infty$, $\sum_{t}\gamma_t^2<\infty$, and $\frac{\gamma_t}{\alpha_t}\rightarrow 0 \text{ for } t\rightarrow +\infty
$. These conditions required for the learning step-sizes guarantee that the policy parameter $\theta_t$ is updated at a slower time scale than that of the function approximation $w_t$, and thus assure the convergence of the learning rule~\cite{konda2000actor,borkar1997stochastic,bertsekas1996neuro}.

\subsubsection{Evolution of cooperative behaviors}
To capture the evolutionary process of cooperation, we first denote the number of $C$ players in the population by $n_t$ at time $t$. Since there is only one individual to alter its action per step in our model, all possible changes of $n_t$ in each time step will be limited to increasing by one, decreasing by one, or keeping unchanged. It implies that the evolutionary process of cooperation can be formulated as a Markov chain $\{n_t\}$ defined over the finite state space $\mathcal{N}=\{0,1,2,\ldots,N\}$. Meanwhile, the transition probability from $n_t=u \in \mathcal{N}$ to $n_{t+1}=v \in \mathcal{N}$ can be calculated by
\begin{equation}\label{eq}
\begin{split}
p_{u,v}(t)&=\sum_{s\in\mathcal{S}}Pr\{s_t=s|s_0,\pi\}
\sum_{j\in\mathcal{J}}
  \left\{
  \begin{array}{ll}
    p_Cp_{C,j}\pi(s,j,C;\theta_t,\beta)+p_Dp_{D,j}
\pi(s,j,D;\theta_t,\beta), & \hbox{for } v=u; \\
    p_Cp_{C,j}\pi(s,j,D;\theta_t,\beta), & \hbox{for } v=u-1;\\
    p_Dp_{D,j}\pi(s,j,C;\theta_t,\beta), & \hbox{for } v=u+1; \\
    0, & \hbox{otherwise};
  \end{array}
\right.
\end{split}
\end{equation}
where $p_C=u/N$ (resp. $p_D=(N-u)/N$) is the probability that an individual who previously took action $C$ (resp. $D$) is chosen as the focal player at time $t$; $p_{C,j}$ (resp. $p_{D,j}$) is the average probability that players who previously took action $C$ (resp. $D$) encounter $j$ opponents taking action $C$ among $d-1$ co-players at time $t$. It is clear that the Markov chain is non-stationary because the transition probabilities change with time.
\par
To find the average abundance of cooperators in the population, we first note that the actor-critic reinforcement learning converges~\cite{sutton2000policy,konda2000actor} and the environmental dynamics have been described by an irreducible and aperiodic Markov chain. That is, the policy parameter $\theta_t$ will converge to a local optimum of $\rho(\pi)$, $\theta^*=\lim_{t\rightarrow \infty}\theta_t$ (see Supporting Information SI.1 for details), and the dynamics of environmental states will have a unique stationary distribution $d^\pi(s)=\lim_{t\rightarrow \infty}Pr\{s_t=s|s_0,\pi\}$. Using these two facts, it follows that the probability transition matrix $P(t)=[p_{u,v}(t)]_{(N+1)\times (N+1)}$ will converge to $P^*=[p_{u,v}^*]_{(N+1)\times (N+1)}$ for $t\rightarrow\infty$,
where
\begin{equation}\label{eq9}
\begin{split}
p_{u,v}^*&=\lim_{t\rightarrow\infty}p_{u,v}(t) \\
&=\sum_{s\in\mathcal{S}}d^\pi(s)\sum_{j\in\mathcal{J}}
  \left\{
  \begin{array}{ll}
    p_Cp_{C, j}\pi(s,j,C;\theta^*,\beta)+p_Dp_{D, j}
\pi(s,j,D;\theta^*,\beta), & \hbox{for } v=u; \\
    p_Cp_{C, j}\pi(s,j,D;\theta^*,\beta), & \hbox{for } v=u-1;\\
    p_Dp_{D, j}\pi(s,j,C;\theta^*,\beta), & \hbox{for } v=u+1; \\
    0, & \hbox{otherwise}.
  \end{array}
\right.
\end{split}
\end{equation}
In addition, it is noteworthy that the Markov chain described by the probability transition matrix $P^*$ will be irreducible and aperiodic. This is because based on the probability transition matrix $P^*$, any two states of the Markov chain are accessible to each other and the period of all states is $1$. Hence, one can conclude that the non-stationary Markov chain $\{n_t\}$ is strongly ergodic~\cite{isaacson1976markov,bowerman1974nonstationary} and there exists a unique long-run (i.e., stationary) distribution $X=[x_n]_{1\times (N+1)}, n\in\mathcal{N}$. Therein, the distribution $X$ can be obtained by calculating the left eigenvector corresponding to eigenvalue $1$ of the probability transition matrix $P^*$, i.e., the unique solution to $X(P^*-I)=\mathbf{0}_{N+1}$ and $\sum_{n\in\mathcal{N}}x_n=1$, where $I$ is the identity matrix with the same dimension of $P^*$ and $\mathbf{0}_{N+1}$ is the row vector with $N+1$ zero entries. When the system has reached the stationary state, the average abundance of $C$ players in the population can be computed by $\langle x_C\rangle=\sum_{n\in \mathcal{N}} (x_n\cdot n/N)$. If $\langle x_C\rangle>1/2$, it implies that $C$ players are more abundant than $D$ players in the population.

\section{Results}

\subsection{Conditions for the prevalence of cooperation}
We first study the condition under which cooperation can be favored over defection, and restrict our analysis in the limit of weak selection ($\beta\rightarrow 0$) given that finding a closed-form solution to this problem for arbitrary selection intensity is usually NP-complete or \# P-complete~\cite{ibsen2015computational}. In the absence of mutations, such a condition can be obtained in general by comparing the fixation probability of cooperation with that of defection~\cite{nowak2004emergence}. In our model, however, how players update their actions is conducted by the policy with an exploration-exploitation tradeoff, which possesses a similar property as the mutation-selection process~\cite{tuyls2003selection}. Thus, in this case, we need to calculate the average abundance of $C$ players when the population has reached the stationary state, and determine whether it is higher than that of $D$ players~\cite{tarnita2009strategy}.
Using all $\mathrm{a}_j(s)$ to construct the vector $A=[\mathrm{\textbf{a}}(s^1),\mathrm{\textbf{a}}(s^2),\ldots,
\mathrm{\textbf{a}}(s^M)]^T$, and all $\mathrm{b}_j(s)$ to construct the vector $B=[\mathrm{\textbf{b}}(s^1),\mathrm{\textbf{b}}(s^2),\ldots,
\mathrm{\textbf{b}}(s^M)]^T$, where $\mathrm{\textbf{a}}(s^k)=[\mathrm{a}_{0}(s^k),\mathrm{a}_{1}(s^k),\ldots,\mathrm{a}_{d-1}(s^k)]$ and $\mathrm{\textbf{b}}(s^k)=[\mathrm{b}_{d-1}(s^k),\mathrm{b}_{d-2}(s^k),\ldots,\mathrm{b}_{0}(s^k)]$, $k=1,2,\ldots,M$, it follows that under weak selection the average abundance of $C$ players in the stationary state is (see Supporting Information SI.2 for details)
\begin{equation}
\langle x_C\rangle=\frac{1}{2}+\frac{1}{N}
\left[\sum_{s\in\mathcal{S}}d^{\pi}(s)\theta^{*T}\Phi_s
(A-B)\right]\beta+o(\beta),
\end{equation}
and thus it is higher than that of $D$ players if and only if
\begin{equation}\label{eq10}
  \sum_{s\in\mathcal{S}}d^{\pi}(s)\theta^{*T}\Phi_s
(A-B)>0,
\end{equation}
where $\Phi_s$, for $\forall s\in\mathcal{S}$, are some coefficient matrices needed to be calculated for the given population structure and every environmental state $s$, but independent of both $\mathrm{a}_j(s)$ and $\mathrm{b}_j(s)$ for $\forall j\in\mathcal{J}$ and $\forall s\in\mathcal{S}$.
\par
To obtain an explicit formulation of condition~(\ref{eq10}), we further consider two specific population structures, well-mixed populations and structured populations. In the former case, the interactive links of individuals are described by a complete graph, whereas in the latter case, they are described by a regular graph with node degree $d-1$. When the population size is sufficiently large, in the limit of weak selection, we find that condition (\ref{eq10}) in these two populations reduces to an identical closed form (see Supporting Information SI.3 for details),
\begin{equation}\label{eq11}
\sum_{s\in\mathcal{S}}d^{\pi}(s)
\sum_{j=0}^{d-1}{d-1\choose j}\frac{1}{2^{d+1}}
\theta^{*T}\left[\phi_{s,j,C}-\phi_{s,j,D}
\right]>0.
\end{equation}
\par
\begin{figure}[!htb]
  \centering
  \includegraphics[width=\hsize]{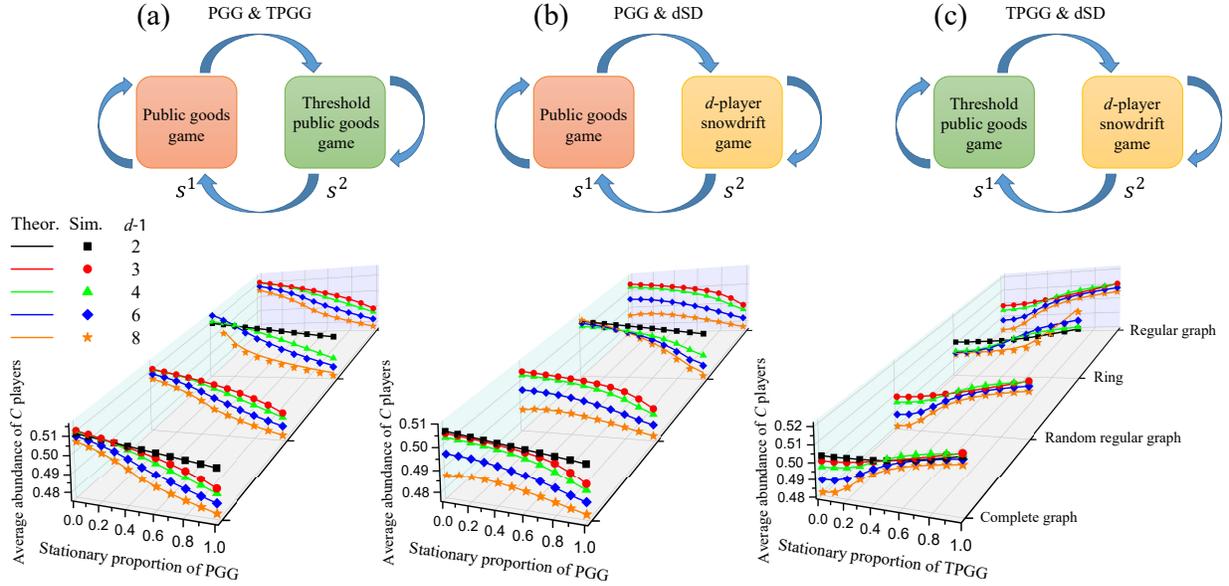}
  \caption{Average abundance of $C$ players in the population as a function of the stationary proportion of different games. In each homogeneous environmental state, $s^1$ or $s^2$, one of the three normal-form games, the public goods game (PGG), threshold public goods game (TPGG), and $d$-player snowdrift game (dSD), is played. In the top row, three transition graphs are depicted to describe how environmental states transit from one to another depending on players' action choices. Corresponding to these three transition graphs, the bottom row shows the average abundance of $C$ players in various population structures, based on numerical calculations and simulations. All simulations are obtained by averaging $40$ network realizations and $10^8$ time steps after a transient time of $10^7$, and $\theta$ is normalized per step to unify the magnitude. Parameter values: $N=400$, $\beta=0.01$, $\mathcal{C}=c=1.0$, $r_{s}=3.0$ in the PGG while $r_{s}=4.0$ in the TPGG, $\mathcal{B}_{s}=12$ in (b) while $\mathcal{B}_{s}=4$ in (c), and $T=[d/2]+1$ ($[\cdot]$ represents the integer part).
  }
  \label{fig2}
\end{figure}

Through extensive agent-based simulations, we validate the effectiveness of this criterion. As illustrated in Fig.~\ref{fig2}, we have calculated the average abundance of $C$ players in the population with two distinct environmental states, $s^1$ and $s^2$, which, for instance, can represent the prosperous state and degraded state of a social-ecological system~\cite{weitz2016oscillating,barfuss2020caring}, respectively. To specify the type of the normal-form multi-player game defined by the payoff Table~\ref{table1} for each given environmental state, in Fig.~\ref{fig2}, we consider that one of the three candidates, the public goods game (PGG)~\cite{hardin1968tragedy}, threshold public goods game (TPGG)~\cite{milinski2008collective,du2014aspiration}, and $d$-player snowdrift game (dSD)~\cite{souza2009evolution}, is played in each state. In these three kinds of games, the implication of defection is unanimous and it means not to contribute. However, in defining cooperative behaviors and calculating payoffs, there are some differences. In the PGG, action $C$ means contributing a fixed amount $c$ to the common pool. After a round of donation, the sum of all contributions from the $d$-player group will be multiplied by a synergy factor $r_s>1$ and then allotted equally among all members, where the value of $r_s$ depends on the current game environment $s$. In this case, the payoffs of cooperators and defectors are computed by $\mathrm{a}_j(s)=(j+1)r_sc/d-c$ and $\mathrm{b}_j(s)=jr_sc/d$, $j\in\mathcal{J}$, respectively. The aforementioned setting is also true for the TPGG, except that there exists a minimum contribution effort, $T$, for players to receive benefits. More specifically, only when the number of $C$ players in the $d$-player game is not smaller than $T$, can each player receive a payoff from the common pool; otherwise, everyone gets nothing. It then follows that a $C$ player will receive a payoff $\mathrm{a}_j(s)=(j+1)cr_s/d-c$ for $j\geq T-1$ and $\mathrm{a}_j(s)=0$ otherwise, whereas a $D$ player will receive $\mathrm{b}_j(s)=jcr_s/d$ for $j\geq T$ and $\mathrm{b}_j(s)=0$ otherwise. Different from the PPG and TPGG, in the dSD, action $C$ means endowing everyone with a fixed payoff $\mathcal{B}_s$ and simultaneously sharing a total cost $\mathcal{C}$ evenly with the other $C$ players, where $\mathcal{B}_s$ depends on the environmental state $s$. In this case, the payoffs of cooperators and defectors are then changed to $\mathrm{a}_j(s)=\mathcal{B}_s-\mathcal{C}/(j+1)$ for $j\in\mathcal{J}$, and $\mathrm{b}_j(s)=\mathcal{B}_s$ for $j>0$ and $\mathrm{b}_0(s)=0$, respectively. As shown in Fig.~\ref{fig2}, the analytical predictions of the average abundance of $C$ players are highly consistent with simulation results, which suggests that criterion (\ref{eq11}) is effective for determining whether cooperation can outperform defection.
\par
Moreover, conditions (\ref{eq10}) and (\ref{eq11}) offer us an intuitional theoretical interpretation of why the environment can mediate social dilemmas~\cite{estrela2019environmentally}. As shown in Fig.~\ref{fig2}, in an identical scenario, the average abundance of $C$ players is always less than $1/2$ in the homogeneous state where the PGG is played, whereas it is greater than $1/2$ in some homogeneous states where a TPGG or dSD is played. The reason is that
the social dilemma in the TPGG and dSD is weaker than that in the PGG. Thus, cooperation in these two kinds of games is easier to evolve. Namely, if the environment is homogeneous, condition (\ref{eq10}) or (\ref{eq11}) in the PGG is more difficult to be satisfied in contrast to the TPGG or dSD. Due to the existence of the underlying transition of the environment, however, the population may have some opportunities to extricate itself from those hostile environmental states where defection is dominant (e.g., the state of the PGG). This case is especially likely after some prosocial behaviors have been implemented by players~\cite{hilbe2018evolution,Su19Evolutionary,barfuss2020caring}. As such, the population will spend some time staying in the states where defection is not always favorable (e.g., the TPGG or dSD). Consequently, the changing environment balances the conditions that favor vs. undermine cooperation, and meanwhile the social dilemma that the population is confronted with is diluted. Such an observation is also in line with the fact that the final outcome of whether cooperation can evolve is a convex combination of those results in each homogeneously environmental state, as shown in conditions (\ref{eq10}) and (\ref{eq11}).

\subsection{Learning vs. non-learning}  
Here, we first exclude the effect of reinforcement learning, and apply our model framework to study two prototypical non-learning updating processes, the smoothed best response~\cite{szabo2007evolutionary} and the aspiration-based update~\cite{du2014aspiration,wu2018individualised}. For the former, in each time step, the focal player chosen in our model revises its action by comparing the payoff of cooperation with that of defection, and the more profitable action will be adopted. Instead of doing this in a deterministic fashion, in many real-life situations, it is more reasonable to assume that the choice of the best response is achieved smoothly and influenced by noise. One typical form to model this process is the Fermi function~\cite{szabo2007evolutionary},
\begin{equation}\label{eq12}
   \pi(s,j,a;\beta)=\frac{1}
  {1+e^{-\beta [\mathcal{R}_{s,j}^a-\mathcal{R}_{s,j}^b]}},\ \forall s\in \mathcal{S}, j\in\mathcal{J}, a,b(\neq a)\in \mathcal{A},
\end{equation}
which specifies the probability for the focal player to choose action $a\in\mathcal{A}$. For the latter, however, the focal player determines whether to switch to a new action by comparing the action's payoff with an internal aspiration level. If the payoff is higher than the aspiration level, the focal player will switch to that action with a higher probability. Otherwise, its action is more likely to keep unchanged. Similarly, the commonly used form to quantify the probability that the focal player switches to the new action $a\in\mathcal{A}$ is still the Fermi function~\cite{du2014aspiration,wu2018individualised},
\begin{equation}\label{eq13}
   \pi(s,j,a;\beta)=\frac{1}
  {1+e^{-\beta [\mathcal{R}_{s,j}^a-\mathcal{E})]}},\ \forall s\in \mathcal{S}, j\in\mathcal{J}, a\in \mathcal{A},
\end{equation}
where a constant aspiration level $\mathcal{E}$ is adopted because heterogenous aspirations~\cite{wu2018individualised} or time-varying aspirations (see Supporting Information SI.4) cannot result in altering the
evolutionary outcome under weak selection. Using these two non-learning updating functions as the decision-making policy of the focal player, under our model framework, we find that in the limit of weak selection, cooperation is more abundant than defection if and only if
\begin{equation}\label{eq14}
  \sum_{s\in\mathcal{S}}d^{\pi}(s)\sum_{j\in\mathcal{J}}
\sigma_j[\mathrm{a}_j(s)-\mathrm{b}_{d-1-j}(s)]>0,
\end{equation}
where $\sigma_j$, $\forall j\in\mathcal{J}$, are some coefficients needed to be calculated for the given population structure, but independent of both $\mathrm{a}_j(s)$ and $\mathrm{b}_j(s)$. In either well-mixed populations or structured populations, we find that the coefficients are $\sigma_j={d-1\choose j}/2^{d+1}$ for the smoothed best response and $\sigma_j={d-1\choose j}/2^{d+2}$ for the aspiration-based update (see Supporting Information SI.4 for details). In particular, if the population consistently stays in a homogeneous environment, condition~(\ref{eq14}) will reduce to the ``sigma-rule'' in the context of multi-player games~\cite{wu2013dynamic}.
\par
\begin{figure}[!htb]
  \centering
  \includegraphics[width=\hsize]{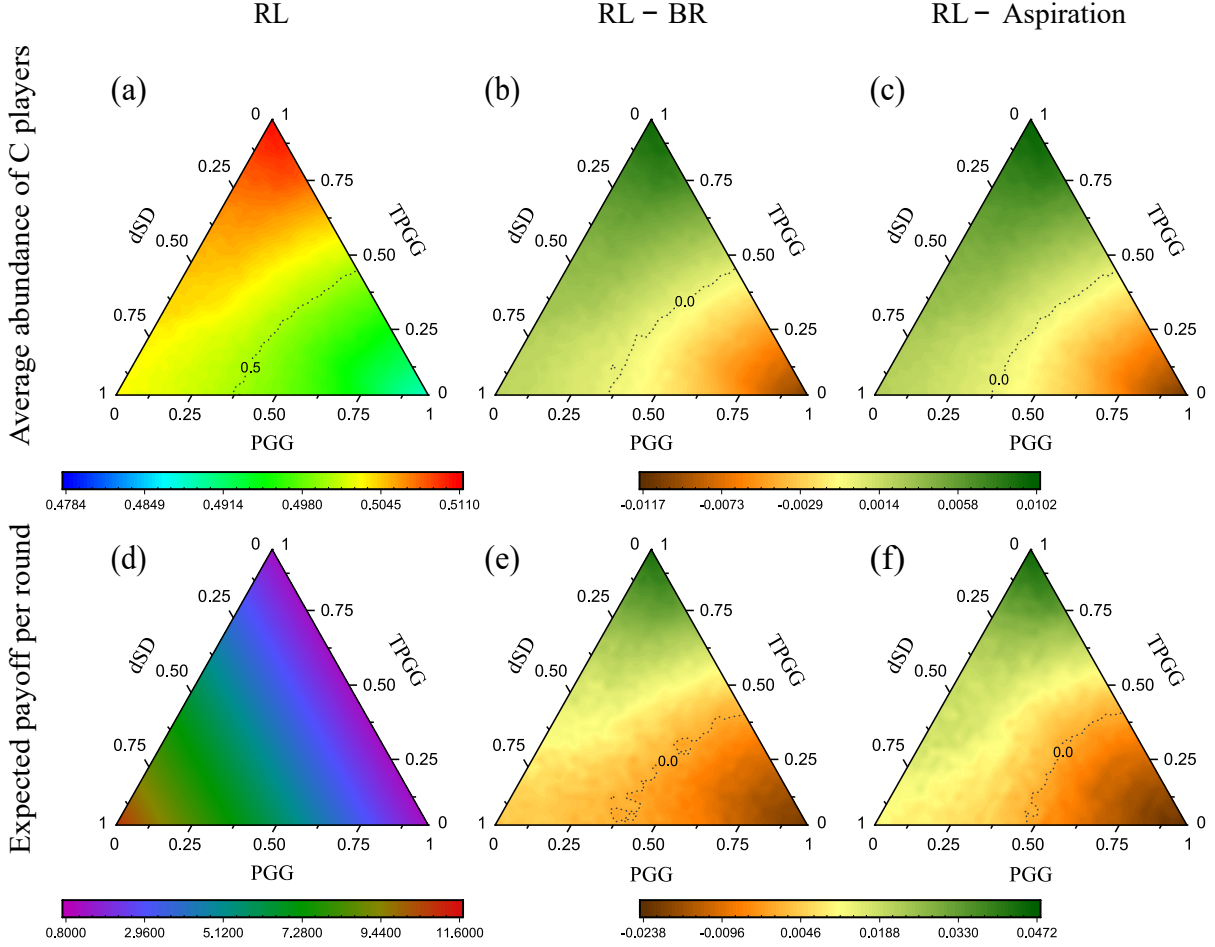}
  \caption{Differences in the average abundance of $C$ players and the expected payoff of players per round between the reinforcement learning (RL) and two non-learning updates. In (a) and (d), we show the average abundance of $C$ players and the expected payoff per round when players update actions via the RL, respectively. Taking them as the benchmark, (b) and (e) illustrate the differences between the RL and the smoothed best response (BR), while (c) and (f) show the gaps between the RL and the aspiration-based rule (Aspiration). The population structure is a lattice network (see Supporting Information Figs.~\ref{figS1}--\ref{figS4} for other population structures with different network degrees). Parameter values: $N=400$, $d=5$, $\beta=0.01$, $\mathcal{C}=c=1.0$, $T=[d/2]+1$, $\mathcal{B}_{s}=12$, $r_{s}=3.0$ for the PGG, and $r_{s}=4.0$ for the TPGG.
  }
  \label{fig3}
\end{figure}

In a population where there are three distinct environmental states and in each state one of the PGG, TPGG, and dSD, is played, we compare the results obtained by learning through reinforcement with those obtained from the two non-learning updates. As illustrated in Fig.~\ref{fig3}, we calculate the average abundance of $C$ players and the expected payoff of focal players per round for all possible stationary distributions of environmental states. Intriguingly, we find that learning enables the adaptation of players to the varying environment. When the population stays in the environment where players are confronted with a weak social dilemma (i.e., the TPGG or dSD will be more likely to be played than the PGG), learning players will have a higher propensity for cooperation than those non-learning players. Moreover, they will reap a higher expected payoff per step. In contrast, when the population stays in the environment where the social dilemma is strong (i.e., the PGG will be more likely to be played than the TPGG and dSD), learning players will have a lower propensity for cooperation and meanwhile they will receive a lower expected payoff per step than non-learning players. Once again, we demonstrate that the analytical results are consistent with the agent-based simulations (see Supporting Information Fig.~\ref{figS5}).

\subsection{Evolutionary dynamics under non-stationary conditions}
The aforementioned analysis mainly focuses on the stationary population environment, i.e., the environmental states have a unique stationary distribution for the long-run evolution. Here, we relax this setup to study the evolutionary dynamics of cooperation under two kinds of non-stationary conditions by agent-based simulations.
\subsubsection{Non-stationary environmental state distribution}
The first case that we are interested in is that the probability distribution of environmental states changes over time. In a population with two environmental states, $s^1$ and $s^2$, we denote the average proportion of the time that the environment stays in state $s^1$ (i.e., the average probability that the environment stays in $s^1$ per step) by $z\in[0,1]$. Then, the average fraction of time in state $s^2$ is $1-z$. To describe the type of the game played in each environmental state, let $s^1$ be the prosperous state where environmental resources are replete and players are at the risk of the tragedy of the commons (i.e., a PGG is played), whereas $s^2$ be the degraded state where environmental resources are gradually depleted. In any state of the environment, cooperation is an altruistic behavior that will increase the common-pool resources, whereas defection is a selfish behavior that will lead the common-pool resources to be consumed. Furthermore, the state of common-pool resources (i.e., the environmental state) will conversely affect individual behaviors. To characterize this feedback relation, we here adopt the difference form of the replicator dynamics with environmental feedbacks~\cite{weitz2016oscillating,tilman2020evolutionary} to describe the evolution of the average time proportion of state $s^1$,
\begin{equation}
  \Delta z(t)=\eta z(t)(1-z(t))(x_C(t)-\bar{x}_C),
\end{equation}
where $\eta$ denotes the positive step-size, $x_C(t)$ is the proportion of $C$ players in the population at time $t$, and $\bar{x}_C$ is the tipping point of the proportion of $C$ players. If the proportion of $C$ players $x_C(t)$ is above the tipping point $\bar{x}_C$, it means that the number of cooperators is competent to sustain the supply of the common-pool resources. At the same time, the environment will also be more likely to stay in the prosperous state $s^1$, leading $z(t)$ to increase. Otherwise, cooperators will be insufficient and the public resources will be continuously consumed. In this case, $z(t)$ will decrease as the environment will more frequently stay in the degraded state $s^2$.
\par
\begin{figure}
  \centering
  \includegraphics[width=\hsize]{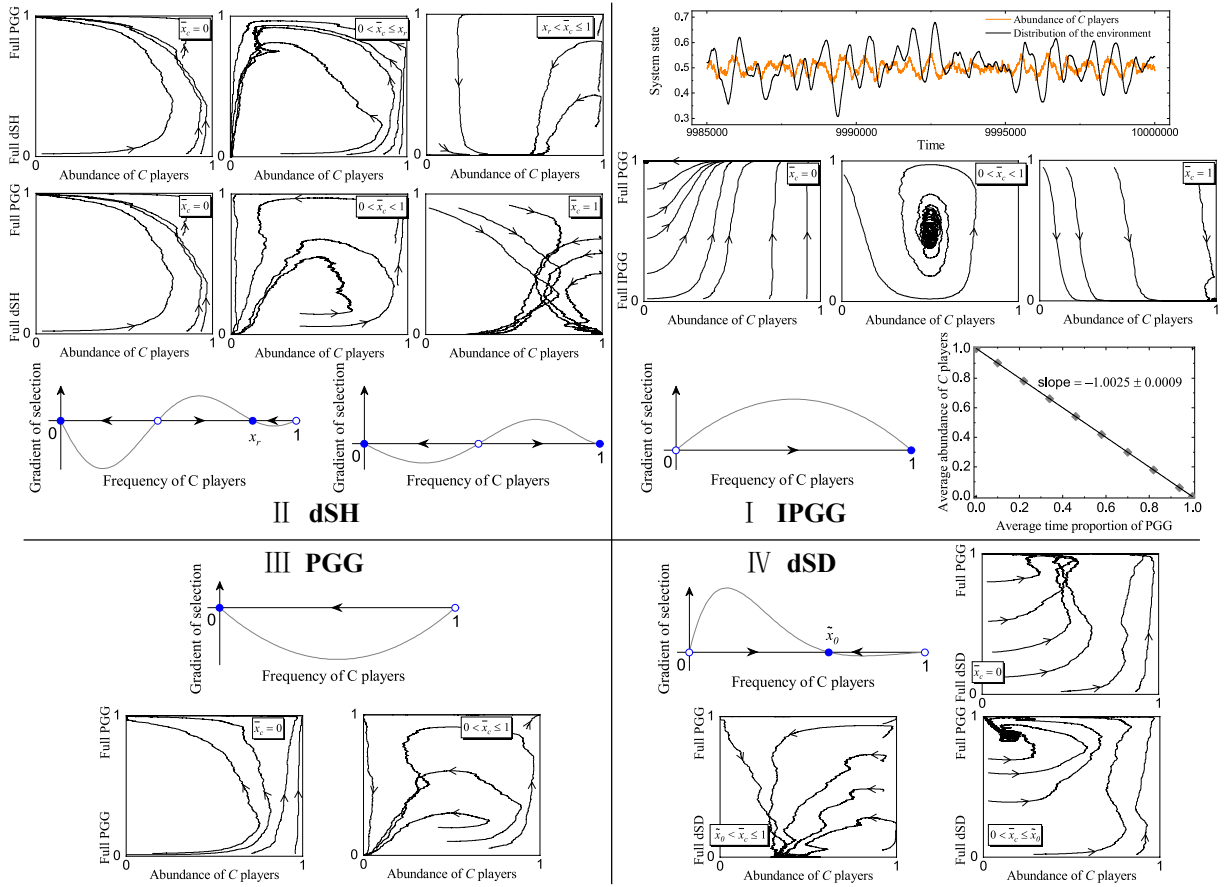}\\
  \caption{Co-evolutionary dynamics of cooperation and the environment under moderate selection intensity. In each panel, sub-figures for the gradient of selection are obtained by replicator dynamics~\cite{hofbauer1998evolutionary,souza2009evolution,pacheco2009evolutionary}. The direction of evolution is indicated by arrows. Blue solid circles are used to depict stable equilibria, while open blue circles are used to depict
  unstable equilibria. From panel~\uppercase\expandafter{\romannumeral1} to panel~\uppercase\expandafter{\romannumeral4}, the IPGG, dSH, PGG, and dSD are used, respectively, to specify the normal-form game played in state $s^2$. In state $s^1$, players participate in a PGG. The phase graphs in each panel show the co-evolutionary dynamics of the average proportion of $C$ players and the time proportion of the PGG for different value intervals of the tipping point $\bar{x}_C$. Corresponding to the value interval $0<\bar{x}_C<1$, the first row in panel \uppercase\expandafter{\romannumeral1} shows the persistent oscillations of cooperation and the environment. The bottom right sub-figure in panel \uppercase\expandafter{\romannumeral1} shows the linear relation between  the average abundance of $C$ players and the average time proportion of the PGG, which suggests that condition~(\ref{eq11}) is still effective for relatively moderate selection intensity to some extent. The first row in panel \uppercase\expandafter{\romannumeral2} corresponds to the case where there is a stable and an unstable interior equilibrium under replicator dynamics (the bottom left), whereas the second row corresponds to the case where there is a unique interior unstable equilibrium (the bottom right). The population structure is finite and well-mixed. Parameter values: $N=400$, $d=5$, $\beta=2$, $\mathcal{C}=c=1.0$, $\mathcal{B}_s=12.0$, $r_s=3.0$ for all panels, except, in panel $\mathrm{II}$, $r_s=4.0$ and $T=3$ for the first row, and $r_s=5.01$ and $T=4.0$ for the second row.
  }\label{fig4}
\end{figure}

In Fig.~\ref{fig4}, we consider that in the prosperous state $s^1$ players play a PGG. However, in the degraded state $s^2$, one of the four different games, the PGG, IPGG (inverse public goods game, which reverses the payoffs of action $C$ and $D$ in the PGG), dSH ($d$-player stag hunt game, which is a variant of the TPGG, and whose only difference from the TPGG is that cooperators always entail a cost $c$ even if $j<T$), and dSD, is played. The reason that we select these four types of games is twofold. On the one hand, they are commonly used to mimic the essence of a vast number of real-life group interactions~\cite{archetti2012game}; on the other hand, they encompass all possible evolutionary behaviors for the frequency-dependent selection between $C$ and $D$ under the classic replicator dynamics~\cite{hofbauer1998evolutionary}: $D$ dominance, $C$ dominance, bistability, and coexistence (see Fig.~\ref{fig4}). Through agent-based simulations, in Fig.~\ref{fig4}, we show the co-evolutionary dynamics of cooperation and environmental states under moderate selection intensity. Depending on the game type and the value of the tipping point $\bar{x}_C$, the population emerges various dynamic behaviors. Particularly, although our model is stochastic and incorporates the effect of environment and learning, we can still observe those dominance, bistability, and coexistence behaviors analogously obtained under the deterministic replicator dynamics. In addition, when replicator dynamics predict that cooperation will be the dominant choice in the degraded state $s^2$, our results show some persistent oscillations between cooperation and the environment (panel~\uppercase\expandafter{\romannumeral1} in Fig.~\ref{fig4}).

\subsubsection{External incentives}
\begin{figure}
  \centering
  \includegraphics[width=\hsize]{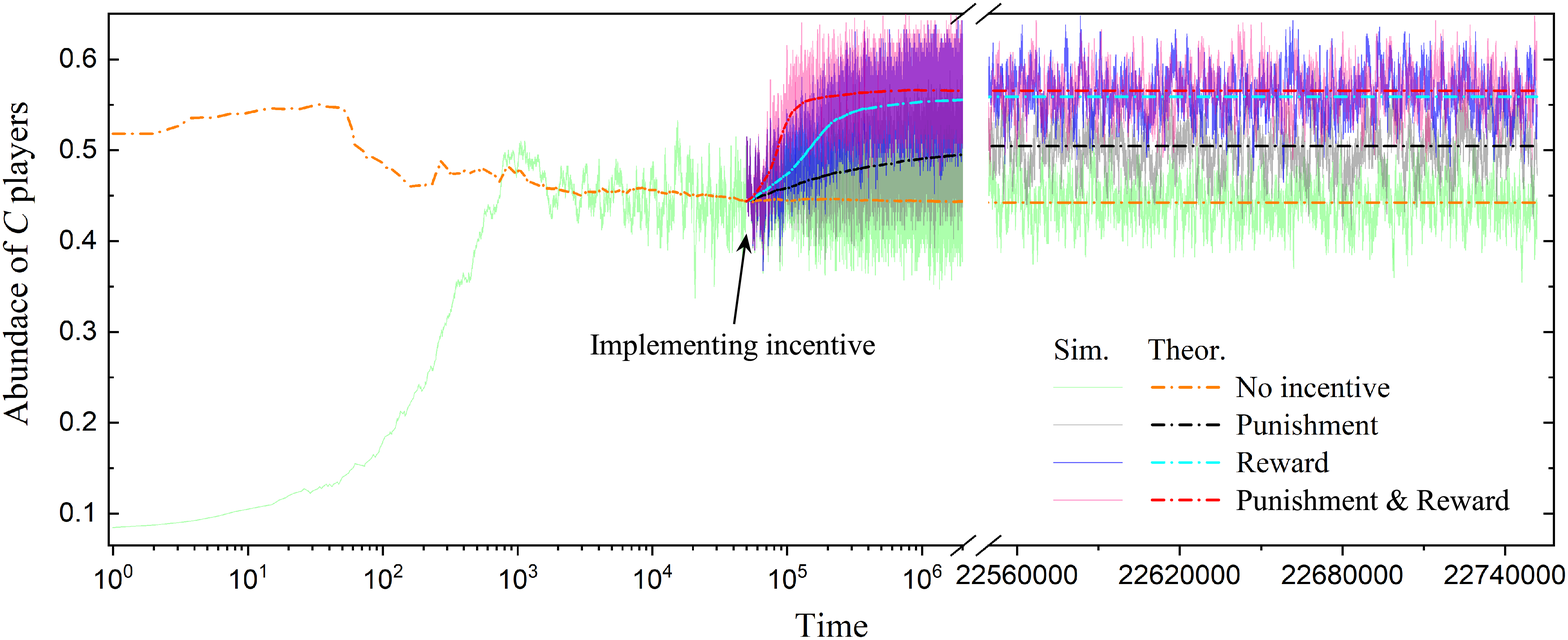}\\
  \caption{Evolution of cooperation under the influence of external incentives. Light solid lines indicate simulations whereas dash dot lines correspond to analytical results. During the evolution, we separately implement punishment and reward, or jointly enforce them to all players in the population with four environmental states, where in each state, one of the IPGG, dSH, PGG, and dSD is played. One can observe that these external incentives markedly enhance the abundance of $C$ players in the population. The population structure is a lattice network. Parameter values: $N=400$, $d=5$, $\beta=0.05$, $\mathcal{C}=c=1.0$ for all games, except $r_{s}=3.0$ for the PGG and TPGG, $r_{s}=5.0$ and $T=[d/2]+2$ for the dSH, and $\mathcal{B}_{s}=12$ for the dSD.
  }\label{fig5}
\end{figure}

Another interesting case is the existence of external incentives, which will undermine the stationarity of the payoff structure of the game. Like two sides of a coin, reward and punishment are two diametrically opposed external incentives for sustaining human cooperation~\cite{fehr2003nature,sigmund2010social,perc2017statistical}. The former is a type of positive incentives where players who cooperate will get an additional benefit, while the latter is a kind of negative incentives where those who defect will be sanctioned and pay a fine. At a certain moment during the evolution of cooperation, we separately implement punishment and reward, or jointly enforce them to all players in the population with four environmental states. We find that both punishment and reward are effective tools in promoting cooperation, even if the game environment may change (see Fig.~\ref{fig5}).

\section{Discussion}
In natural populations, the biotic and abiotic environment where organisms are exposed persists variations in time and space. To win the struggle for survival in this uncertain world, organisms have to timely adjust their behaviors in response to the fluctuation of their living environments~\cite{levins1968evolution,meyers2002fighting}. For the longstanding conundrum of how cooperation can evolve, however, the majority of the existing evolutionary interpretations has been devoted to understanding the static interactive scenarios~\cite{west2007evolutionary,nowak2006five}. Therefore, when individual interactions, especially involving multiple players at a time, occur in the changing environment, determining whether cooperation can evolve will become fairly tricky. Here, we developed a general model framework that incorporates an adaptation mechanism of reinforcement learning to investigate the evolution of cooperative behaviors in the constantly changing multi-player game environment. Our model not only considers the interplay between players' behaviors and environmental variations, but also incorporates a cognitive or psychological feedback loop where players' choices determine the game outcome, and in turn are affected by it. Such a setup is, to some extent, analogous to the human decision in the context of the hybrid human-machine cooperation~\cite{crandall2018cooperating}, a key research theme in the emerging interdisciplinary field -- machine behavior~\cite{rahwan2019machine}, in which humans can use algorithms to make decisions and subsequently the training of the same algorithms is affected by those decisions.
\par
The importance of environmental variations in population dynamics has long been recognized in theoretical ecology and population biology~\cite{macarthur1970species,levins1968evolution,ROSENBERG20201}.
In a realistic social or ecological system, individual behaviors and environmental variations are inevitably coupled together~\cite{macarthur1970species,levins1968evolution}. By consuming, transforming, or producing common-pool resources, for example, organisms are enabled to alter their living environments, and consequently, such modification may consequentially be detrimental or beneficial to their survival~\cite{estrela2019environmentally}. Our analytical condition for determining whether cooperation can be favored over defection indeed provides us a plausible theoretical explanation for this phenomenon. If mutual actions of individuals lead the environment to transit from a preferable state where cooperation is more profitable to a hostile one where defection is more dominant, cooperation will be suppressed. Otherwise, cooperation will flourish. In particular, if the population has access to switching among multiple environmental states, the environment will play the role of intermediates in social interactions and the final outcome of whether cooperation can evolve will be the synthesis of results in each environmental state. Such an observation is different from the recent findings where
game transitions can result in a more favorable outcome for cooperation even if all individual games favor defection~\cite{hilbe2018evolution,Su19Evolutionary}. One important reason for this is that we do not follow the scheme to explicitly assign how the environment depends on individual actions to transit from one state to another, but rather use an ergodic Markov chain to characterize the dynamics of the environment. Thus, in this sense, our model is more general and can be applied to a large variety of environmental transition processes.
\par
Moreover, compared with the existing studies on the evolution of cooperation in the changing environment~\cite{weitz2016oscillating,hilbe2018evolution,tilman2020evolutionary,chen2018punishment,Su19Evolutionary,hauert2019asymmetric},
another striking difference is that, apart from the environmental feedback, our model introduces the learning mechanism of reinforcement. Since, when the environment changes, the previous decision-making scheme adopted by individuals may fail to work, they must learn how to adjust their behaviors in response to the contingencies given by the environment, in order to obtain a higher fitness. Such a scenario is also closely related to some recent work across disciplines, including statistical physics~\cite{sato2005stability,macy2002learning,sato2002chaos,galla2013complex,barfuss2019deterministic}, artificial intelligence~\cite{sutton2018reinforcement,tuyls2003selection,bloembergen2015evolutionary}, evolutionary biology~\cite{dridi2014learning,dridi2018learning}, and neuroscience~\cite{dayan2008reinforcement,khalvati2019modeling}. However, their dominant attention has been paid to learning dynamics, the deterministic limit of the learning process, the design of new learning algorithms in games, or neural computations. In comparison, our model is discrete and stochastic, and focuses on multi-player stochastic games. In particular, our analysis for the game system is systematic and encompasses a variety of factors, such as group interactions, spatial structures, and environmental variations. In addition, our work may offer some new insight into the interface between reinforcement learning and evolutionary game theory from the perspective of approximate solution methods~\cite{sutton2018reinforcement}, because most existing progress in combining tools from these two fields to explore the interaction of multiple agents is based on the tabular solution methods~\cite{sato2005stability,tuyls2003selection,barfuss2019deterministic,bloembergen2015evolutionary}.
\par
In the present work, one of the main limitations is that the strategic update is restricted to the asynchronous type and the learning experience is required to be shared across individuals. Although such a setup is appropriate in those scenarios where individuals modify their strategies independently, and typical in economics applications and for overlapping generations~\cite{szabo2007evolutionary}, it has been suggested that the unanimous satisfactory decisions reached by all individuals based on asynchronous updates cannot always be guaranteed by synchronous updates~\cite{ramazi2016networks}. In particular, if individuals are able to communicate with each other via a network or leverage the perceived information to model
and infer the choices of others~\cite{bu2008comprehensive,camerer2011behavioral,khalvati2019modeling}, the asynchronous update will become more problematic. Thus, further work on synchronously strategic revisions is worthy of exploring in the future. Of course, such an extension will also be full of challenges, because updating strategies simultaneously for multiple agents will inevitably give rise to some intractable problems, such as the curse of dimensionality, requirement for coordination, nonstationarity, and exploration-exploitation tradeoff~\cite{bu2008comprehensive}. Moreover, some further efforts should be invested in the partial observability of the Markov environmental states and relaxing the perfect environmental information required in our model to the local or unpredictable type~\cite{kaelbling1998planning}.

\section*{Acknowledgments}
This work was supported by the National Natural
Science Foundation of China (Grant 61751301 and Grant 61533001). F. Huang acknowledges the support
from China Scholarship Council (Grant 201906010075). The work of Cao was supported in part by the European Research Council (ERC-CoG-771687) and the Netherlands Organization for Scientific Research (NWO-vidi-14134).

\newpage

\begin{center}
\Large\textbf{Supporting Information}
\end{center}

\setcounter{equation}{0}
\setcounter{figure}{0}
\setcounter{section}{0}
\setcounter{table}{0}
\renewcommand{\thesection}{SI.\arabic{section}}
\renewcommand{\thesubsection}{SI.\arabic{section}.\arabic{subsection}}
\renewcommand{\theequation}{SI.\arabic{equation}}
\renewcommand{\thetable}{SI.\arabic{table}}
\renewcommand\thefigure{S\arabic{figure}}
\setcounter{figure}{0}

\section{Algorithm derivation for the actor-critic reinforcement learning}
Here, we derive the algorithm of the actor-critic reinforcement learning adopted in our model, using the method proposed in Refs.~\cite{sutton2000policy,konda2000actor}. First, we define the state value function $V^{\pi}(s,j)$ under the policy $\pi$ for a given state pair, $s$ and $j$, by $V^{\pi}(s,j)\triangleq
\sum_{a\in\mathcal{A}}\pi(s,j,a;\theta,\beta)Q^{\pi}(s,j,a)$, $\forall s\in\mathcal{S}$ and $j\in\mathcal{J}$, and let $\rho(\pi)$ be the performance measure of policy $\pi$ with respect to the policy parameter $\theta$. The goal of the actor-critic reinforcement learning is to seek to maximize the performance. Thus, the policy parameter is updated in the direction of the gradient ascent of $\rho(\pi)$,
\begin{equation}\label{eqa1}
  \theta_{t+1}=\theta_{t}+\gamma_t\frac{\partial \rho(\pi)}{\partial \theta_t},
\end{equation}
where $\gamma_t$ is the positive step size. It is clear that if this iteration can be achieved, $\theta_t$ will be assured to converge to the local optimum of $\rho(\pi)$. In the following, we proceed to derive an unbiased estimator of the gradient $\frac{\partial \rho(\pi)}{\partial \theta}$.
\par
Using the definition of $Q^\pi(s,j,a)$, we first have
\begin{equation}
  \begin{split}
    Q^\pi(s,j,a)&=\sum_{t=1}^{\infty}\mathbb{E}\{r_t-\rho(\pi)
|s_0=s,j_0=j,a_0=a,\pi\} \\
&=\sum_{s'\in\mathcal{S},j'\in\mathcal{J}}Pr(s',j'|s,j,a)[
\mathcal{R}_{s,j}^a-\rho(\pi)  \\
&+\sum_{a\in\mathcal{A}}\pi(s',j',a;\theta,\beta)\sum_{t=1}^{\infty}\mathbb{E}\{r_t-\rho(\pi)
|s_0=s',j_0=j',a_0=a,\pi\}]  \\
&=\mathcal{R}_{s,j}^a-\rho(\pi)+\sum_{s'\in\mathcal{S},j'\in\mathcal{J}}Pr(s',j'|s,j,a)\sum_{a\in\mathcal{A}}\pi(s',j',a;\theta,\beta)Q^\pi(s',j',a) \\
&=\mathcal{R}_{s,j}^a-\rho(\pi)+\sum_{s'\in\mathcal{S},j'\in\mathcal{J}}Pr(s',j'|s,j,a)V^{\pi}(s',j'),
  \end{split}
\end{equation}
where $Pr(s',j'|s,j,a)$ is the probability that executing action $a\in\mathcal{A}$ leads the current state pair $(s,j)$ to transit to $(s',j')$ in the next time. Then, the derivative of $V^{\pi}(s,j)$ with respect to $\theta$ can be calculated by
\begin{equation}
\begin{split}
\frac{\partial V^{\pi}(s,j)}{\partial \theta}&=\frac{\partial}{\partial \theta}\sum_{a\in\mathcal{A}}\pi(s,j,a;\theta,\beta)Q^{\pi}(s,j,a),\\
&=\sum_{a\in\mathcal{A}}\left[\frac{\partial \pi(s,j,a;\theta,\beta)}{\partial \theta}Q^{\pi}(s,j,a)+\pi(s,j,a;\theta,\beta)
\frac{\partial Q^{\pi}(s,j,a)}{\partial \theta}\right]\\
&=\sum_{a\in\mathcal{A}}\left[\frac{\partial \pi(s,j,a;\theta,\beta)}{\partial \theta}Q^{\pi}(s,j,a)+\right.\\
&\left.\pi(s,j,a;\theta,\beta)
\frac{\partial }{\partial \theta}\left(\mathcal{R}_{s,j}^a-\rho(\pi)+\sum_{s'\in\mathcal{S},j'\in\mathcal{J}}
Pr(s',j'|s,j,a)V^{\pi}(s',j')\right)\right]\\
&=\sum_{a\in\mathcal{A}}\left[\frac{\partial \pi(s,j,a;\theta,\beta)}{\partial \theta}Q^{\pi}(s,j,a)+\right.\\
&\left.\pi(s,j,a;\theta,\beta)
\left(-\frac{\partial \rho(\pi)}{\partial \theta}+
\sum_{s'\in\mathcal{S},j'\in\mathcal{J}}Pr(s',j'|s,j,a)\frac{\partial V^{\pi}(s',j')}{\partial \theta}\right)\right].
\end{split}
\end{equation}
Therefore, it leads to
\begin{equation}
\begin{split}
  \frac{\partial \rho(\pi)}{\partial \theta}&=\sum_{a\in\mathcal{A}}\left[\frac{\partial \pi(s,j,a;\theta,\beta)}{\partial \theta}Q^{\pi}(s,j,a)+\pi(s,j,a;\theta,\beta)  \sum_{s'\in\mathcal{S},j'\in\mathcal{J}}Pr(s',j'|s,j,a)\frac{\partial V^{\pi}(s',j')}{\partial \theta}\right]\\
&-\frac{\partial V^\pi(s,j)}{\partial \theta}.
\end{split}
\end{equation}
Multiplying both sides of the equation by $d^\pi(s)p_{\cdot j}$ and summing over $s\in\mathcal{S}$ and $j\in\mathcal{J}$ yield
\begin{equation}\label{eqa5}
\begin{split}
\sum_{s\in\mathcal{S}}d^\pi(s)\sum_{j\in\mathcal{J}}p_{\cdot j}  \frac{\partial \rho(\pi)}{\partial \theta}&= \sum_{s\in\mathcal{S}}d^\pi(s)\sum_{j\in\mathcal{J}}p_{\cdot j}
\sum_{a\in\mathcal{A}}\frac{\partial \pi(s,j,a;\theta,\beta)}{\partial \theta}Q^{\pi}(s,j,a) \\
&+\sum_{s\in\mathcal{S}}d^\pi(s)\sum_{j\in\mathcal{J}}p_{\cdot j}
\sum_{a\in\mathcal{A}}\pi(s,j,a;\theta,\beta)
\sum_{s'\in\mathcal{S},j'\in\mathcal{J}}Pr(s',j'|s,j,a)\frac{\partial V^{\pi}(s',j')}{\partial \theta} \\
&-\sum_{s\in\mathcal{S}}d^\pi(s)\sum_{j\in\mathcal{J}}p_{\cdot j}\frac{\partial V^\pi(s,j)}{\partial \theta}.
\end{split}
\end{equation}
In addition, note that $\sum_{s\in\mathcal{S}}d^\pi(s)\sum_{j\in\mathcal{J}}p_{\cdot j}=1$ and $\sum_{s\in\mathcal{S}}d^\pi(s)\sum_{j\in\mathcal{J}}p_{\cdot j}
\sum_{a\in\mathcal{A}}\pi(s,j,a;\theta,\beta) \\ Pr(s',j'|s,j,a)=Pr(s',j')$, where $Pr(s',j')$ is the joint probability that the environmental state is $s'$ and the focal player will, on average, encounter $j'$ opponents taking action $C$ among $d-1$ co-players in the stationary state. Further, since $s'$ and $j'$ are independent, we have $Pr(s',j')=d^\pi(s')p_{\cdot j'}$. It follows that Eq.~(\ref{eqa5}) can be rewritten as
\begin{equation}\label{eqa6}
\begin{split}
 \frac{\partial \rho(\pi)}{\partial \theta}&=\sum_{s\in\mathcal{S}}d^\pi(s)\sum_{j\in\mathcal{J}}p_{\cdot j}
\sum_{a\in\mathcal{A}}\frac{\partial \pi(s,j,a;\theta,\beta)}{\partial \theta}Q^{\pi}(s,j,a)\\
&+
\sum_{s'\in\mathcal{S},j'\in\mathcal{J}}d^\pi(s')p_{\cdot j'}\frac{\partial V^{\pi}(s',j')}{\partial \theta}
-\sum_{s\in\mathcal{S}}d^\pi(s)\sum_{j\in\mathcal{J}}p_{\cdot j}\frac{\partial V^\pi(s,j)}{\partial \theta}\\
&=\sum_{s\in\mathcal{S}}d^\pi(s)\sum_{j\in\mathcal{J}}p_{\cdot j}
\sum_{a\in\mathcal{A}}\frac{\partial \pi(s,j,a;\theta,\beta)}{\partial \theta}Q^{\pi}(s,j,a)\\
&=\sum_{s\in\mathcal{S}}d^\pi(s)\sum_{j\in\mathcal{J}}p_{\cdot j}
\sum_{a\in\mathcal{A}}\pi(s,j,a;\theta,\beta)
\frac{\nabla_{\theta} \pi(s,j,a;\theta,\beta)}{\pi(s,j,a;\theta,\beta)}Q^{\pi}(s,j,a)\\
&=\mathbb{E}_{\pi}\left[\frac{\nabla_{\theta} \pi(s,j,a;\theta,\beta)}{\pi(s,j,a;\theta,\beta)}Q^{\pi}(s,j,a)\right],
\end{split}
\end{equation}
where $\mathbb{E}_{\pi}(\cdot)$ represents the expectation under the policy $\pi$, and $\nabla_{\theta} \triangleq\frac{\partial}{\partial \theta}$. Hence,  Eq.~(\ref{eqa6}) gives an unbiased estimator of $\frac{\partial \rho(\pi)}{\partial \theta}$.
\par
From Eq.~(\ref{eqa6}), we know that the unbiased estimator of $\frac{\partial \rho(\pi)}{\partial \theta}$ depends on $Q^{\pi}(s,j,a)$. However, an exact calculation of $Q^{\pi}(s,j,a)$ is usually impossible. One effective way to deal with this problem is to find a good approximation of this value function~\cite{sutton2018reinforcement}. Let $f_w(s,j,a):\mathcal{S}\times\mathcal{J}\times\mathcal{A}\rightarrow \mathbb{R}$ be the approximation to $Q^\pi(s,j,a)$, with the parameter vector $w\in\mathbb{R}^L$. To obtain $f_w(s,j,a)$, it is natural to update $w$ under the policy $\pi$ by the least square method,
\begin{equation}\label{eqa7}
\begin{split}
  \Delta w_t&\propto -\frac{\partial \parallel \hat{Q}^{\pi}(s,j,a)-f_{w_t}(s,j,a)\parallel_{\pi}^2}{\partial w_t},\\
&\propto\sum_{s\in\mathcal{S}}d^\pi(s)\sum_{j\in\mathcal{J}}p_{\cdot j}
\sum_{a\in\mathcal{A}}\pi(s,j,a;\theta,\beta)[\hat{Q}^{\pi}(s,j,a)-f_{w_t}(s,j,a)]
\nabla_{w_t} f_{w_t}(s,j,a),\\
&\propto \mathbb{E}_{\pi}\left\{[\hat{Q}^{\pi}(s,j,a)-f_{w_t}(s,j,a)]
\nabla_{w_t} f_{w_t}(s,j,a)\right\},
\end{split}
\end{equation}
where ``$\propto$'' is the proportional symbol, $\parallel \hat{Q}^{\pi}(s,j,a)-f_{w_t}(s,j,a)\parallel_{\pi}^2$ defines the distance using the norm $\|Q(s,j,a)\|_{\pi}^2=\sum_{s\in\mathcal{S}}d^{\pi}(s)
\sum_{j\in\mathcal{J}}p_{\cdot j}
\sum_{a\in\mathcal{A}}\pi(s,j,a;\theta,\beta)[Q(s,j,a)]^2$, $\hat{Q}^{\pi}(s,j,a)$ is the unbiased estimator of $Q^{\pi}(s,j,a)$, and $\nabla_{w_t}\triangleq\frac{\partial}{\partial w_t}$. When this iterative process has converged to a local optimum, we have
\begin{equation}\label{eqa8}
\mathbb{E}_{\pi}\left\{[Q^{\pi}(s,j,a)-f_{w}(s,j,a)]
\nabla_w f_{w}(s,j,a)\right\}=0.
\end{equation}
In our model, since $f_w(s,j,a)$ is given in a linear form of features and satisfies the canonical compatible condition~\cite{sutton2000policy} $\nabla_w f_w(s,j,a)=\frac{\nabla_{\theta} \pi(s,j,a;\theta,\beta)}{\pi(s,j,a;\theta,\beta)} $ (see Eq.~(\ref{eq6})), subtracting Eq.~(\ref{eqa8}) from Eq.~(\ref{eqa6}) yields
\begin{equation}\label{eqa9}
\begin{split}
 \frac{\partial \rho(\pi)}{\partial \theta}
&=\mathbb{E}_{\pi}\left[\frac{\nabla_{\theta} \pi(s,j,a;\theta,\beta)}{\pi(s,j,a;\theta,\beta)}Q^{\pi}(s,j,a)\right]
-\mathbb{E}_{\pi}\left\{[Q^{\pi}(s,j,a)-f_{w}(s,j,a)]
\nabla_w f_{w}(s,j,a)\right\},\\
&=\mathbb{E}_{\pi}\left[\frac{\nabla_{\theta} \pi(s,j,a;\theta,\beta)}
{\pi(s,j,a;\theta,\beta)}f_{w}(s,j,a)\right].
\end{split}
\end{equation}
Next, based on the temporal-difference learning~\cite{sutton2018reinforcement}, Eqs.~(\ref{eqa1}) and (\ref{eqa7}) can be written as Eq.~(\ref{eq7}). Particularly, when the conditions required for the learning step-sizes in Eq.~(\ref{eq7}) are satisfied, the algorithm is able to be guaranteed to converge to a local optimum of $\rho(\pi)$ by applying the stochastic approximation theorem~\cite{borkar1997stochastic,bertsekas1996neuro}.

\section{Deriving the condition for cooperation to be favored}
In this section, we derive the condition under which the average abundance of $C$ players, $\langle x_C\rangle$, is more abundant than that of $D$ players in the limit of weak selection, when the population has reached the stationary state. Mathematically, this problem is equivalent to find the condition for $\langle x_C\rangle=\sum_{n\in \mathcal{N}} (x_n\cdot n/N)>1/2$ to be true when $\beta\rightarrow 0$, where $x_n$ is the stationary probability that there are $n$ players of $C$ in the population. As mentioned in Methods, to obtain $x_n$, we need to calculate the stationary distribution $X=[x_n]_{1\times (N+1)},n\in \mathcal{N}$, which is the unique solution to $X(P^*-I)=\mathbf{0}_{N+1}$ and $\sum_{n\in\mathcal{N}}x_n=1$, where $P^*=[p_{u,v}^*]_{(N+1)\times (N+1)}$ is the probability transition matrix. From this equation, we know that each $x_n$ will be a rational polynomial function of $p^{*}_{u,v}$, $u,v\in\mathcal{N}$. In addition, we note that $p^{*}_{u,v}$ is differentiable at $\beta=0$ based on Eqs.~(\ref{eq1}) and (\ref{eq9}). It follows that the stationary probability $x_n$ is differentiable at $\beta=0$. In this case, we rewrite $x_n$ in terms of the first-order Taylor expansion under weak selection $\beta\rightarrow 0$,  $x_n=x_n(0)+x_n'(0)\cdot\beta+o(\beta)$, where $x_n'(0)=\frac{\partial x_n(\beta)}{\partial \beta}|_{\beta=0}$. Substituting $x_n$ into $\langle x_C \rangle$, we then obtain
\begin{equation}\label{eqb1}
  \langle x_C\rangle=\frac{1}{N}\sum_{n\in \mathcal{N}}nx_n(0)+\beta\frac{1}{N}\sum_{n\in \mathcal{N}}nx_n'(0)+o(\beta).
\end{equation}
Particularly, we note that the first term on the right-hand side of this equation is in fact the average abundance of $C$ players in the population when selection is neutral, i.e., $\beta=0$. Thus, in the following, we first prove that in the case of neutral selection $\beta=0$, the average abundance of $C$ players is one half, i.e., $\frac{1}{N}\sum_{n\in \mathcal{N}}nx_n(0)=1/2$. Subsequently, we prove that $x_n'(0)$ will be a linear combination of $\mathrm{a}_j(s)$ and $\mathrm{b}_j(s)$, $\forall j\in\mathcal{J}$ and $\forall s\in\mathcal{S}$. Finally, substituting them into~(\ref{eqb1}), we obtain the condition for cooperation to be favored.

\subsection{The average abundance of $C$ players under neutral selection}
When selection intensity is neutral, i.e., $\beta=0$, we denote the strategic state of individual $i$ in the population by $q_i$, where $i$ belongs to the set $\{1,2,\ldots,N\}$ whose elements are the labels of individuals. If individual $i$ takes action $C$, we assign $q_i=1$, and otherwise $q_i=0$. Then, the total number of $C$ players in the population can be computed by $\sum_{i=1}^{N}q_i$. Since the policy used by the focal player to determine whether to cooperate or defect is a probability distribution function over actions, $q_i$ is a random variable. In particular, when selection is neutral $\beta=0$, every individual chosen as the focal player will take action $C$ or $D$ at random, because $\pi(s,j,C;\theta,0)=\pi(s,j,D;\theta,0)=1/2$ for any $s\in\mathcal{S}$, $j\in\mathcal{J}$, and $\theta\in \mathbb{R}^L$. Thus, the expectation of $q_i$, $\mathbb{E}(q_i)$, will always be $1/2$. As a consequence, the average abundance of $C$ players in the population, $\sum_{i=1}^N\mathbb{E}(q_i)/N$, will be one half. This is equivalent to say
\begin{equation}\label{eqb2}
\frac{1}{N}\sum_{n\in \mathcal{N}}nx_n(0)=1/2.
\end{equation}

\subsection{The linear relation}
We proceed to prove that $x_n'(0)$ can be written by a linear combination of $\mathrm{a}_j(s)$ and $\mathrm{b}_j(s)$, $\forall j\in\mathcal{J}$ and $\forall s\in\mathcal{S}$, with a constant term. First, we rewrite the policy $\pi(s,j,a;\theta,\beta)$ in terms of the first-order Taylor expansion under weak selection, given by
\begin{equation}\label{eqb3}
  \begin{split}
    \pi(s,j,a;\theta,\beta)&=\pi(s,j,a;\theta,0)+
    \frac{\partial \pi(s,j,a;\theta,\beta)}{\partial \beta}|_{\beta=0}\cdot\beta+o(\beta)\\
    &=1/2+\frac{|\mathcal{A}|\theta^T\phi_{s,j,a}-
    \sum_{b\in\mathcal{A}}\theta^T\phi_{s,j,b}}{|\mathcal{A}|^2}
    \beta+o(\beta),
  \end{split}
\end{equation}
where $|\mathcal{A}|$ represents the cardinality of action set $\mathcal{A}$. Next, we define an error function $e_{w}(s,j,a)=Q^{\pi}(s,j,a)-f_{w}(s,j,a)$. In this way, we have $f_w(s,j,a)=Q^{\pi}(s,j,a)-e_{w}(s,j,a)$. In addition, based on Eqs.~(\ref{eq3}), (\ref{eq4}), and (\ref{eq5}), we note that
$Q^{\pi}(s,j,a)$ can be written as a linear combination of $\mathrm{a}_{j}(s)$ and $\mathrm{b}_{j}(s)$, $j\in\mathcal{J}$ and $s\in\mathcal{S}$, without containing any constant terms. Thus, for every $Q$-value function $Q^{\pi}(s,j,a)$, there will always exist a set of coefficients, $\mu_{sl}(s,j,a)$ and $\nu_{sl}(s,j,a)$, such that $Q^{\pi}(s,j,a)=\sum_{s'\in\mathcal{S},j'\in\mathcal{J}}
[\mu_{s'j'}(s,j,a)\mathrm{a}_{j'}(s')+\nu_{s'j'}(s,j,a)\mathrm{b}_{j'}(s')]$.
Substituting it into $f_w(s,j,a)=Q^{\pi}(s,j,a)-e_{w}(s,j,a)$, we have $f_w(s,j,a)=\sum_{s'\in\mathcal{S},j'\in\mathcal{J}}
[\mu_{s'j'}(s,j,a)\mathrm{a}_{j'}(s')+\nu_{s'j'}(s,j,a)\mathrm{b}_{j'}(s')]
-e_{w}(s,j,a)$. On the other hand, based on Eq.~(\ref{eq6}), we can get $ f_w(s,j,a)=w^T[\phi_{s,j,a}
-\sum_{b\in\mathcal{A}}\pi(s,j,b;\theta,\beta)\phi_{s,j,b}]\beta$. It follows that
\begin{equation}\label{}
  \begin{split}
    &w^T[\phi_{s,j,a}
-\sum_{b\in\mathcal{A}}\pi(s,j,b;\theta,\beta)\phi_{s,j,b}]\beta \\
&=\sum_{s'\in\mathcal{S},j'\in\mathcal{J}}
[\mu_{s'j'}(s,j,a)\mathrm{a}_{j'}(s')+\nu_{s'j'}(s,j,a)\mathrm{b}_{j'}(s')]
-e_{w}(s,j,a),
  \end{split}
\end{equation}
for any $w\in\mathbb{R}^L$.
\par
From this equation, we can find that the term on the left-hand side is a linear combination of $\phi_{s,j,C}$ and $\phi_{s,j,D}$ without constant terms, whereas that on the right-hand side is a linear combination of
$\mathrm{a}_{j}(s)$ and $\mathrm{b}_{j}(s)$, $j\in\mathcal{J}$ and $s\in\mathcal{S}$, with a constant term $e_{w}(s,j,a)$. It implies that every element of vector $\phi_{s,j,a}$, $a\in\mathcal{A}$, is able to be written as a linear combination of $\mathrm{a}_{j}(s)$ and $\mathrm{b}_{j}(s)$, $j\in\mathcal{J}$ and $s\in\mathcal{S}$, with a constant term proportional to $e_{w}(s,j,a)$. Thus, $\frac{|\mathcal{A}|\phi_{s,j,a}-
\sum_{b\in\mathcal{A}}\phi_{s,j,b}}{|\mathcal{A}|^2}$ will be a vector whose each element is a linear combination of $\mathrm{a}_j(s)$ and $\mathrm{b}_j(s)$
with a constant term proportional to $e_{w}(s,j,a)$. In particular, we use all $\mathrm{a}_j(s)$ to construct the vector $A=[\mathrm{\textbf{a}}(s^1),\mathrm{\textbf{a}}(s^2),\ldots,
\mathrm{\textbf{a}}(s^M)]^T$, and all $\mathrm{b}_j(s)$ to construct the vector $B=[\mathrm{\textbf{b}}(s^1),\mathrm{\textbf{b}}(s^2),\ldots,
\mathrm{\textbf{b}}(s^M)]^T$, where $\mathrm{\textbf{a}}(s^k)=[\mathrm{a}_{0}(s^k),\mathrm{a}_{1}(s^k),\ldots,\mathrm{a}_{d-1}(s^k)]$ and $\mathrm{\textbf{b}}(s^k)=[\mathrm{b}_{d-1}(s^k),\mathrm{b}_{d-2}(s^k),\ldots,\mathrm{b}_{0}(s^k)]$, $k=1,2,\ldots,M$. Then, there will exist two coefficient matrixes, $\mathbf{U}_{s,j,a}$ and $\mathbf{V}_{s,j,a}$, and a constant vector, $F_{s,j,a}$, such that $\frac{|\mathcal{A}|\phi_{s,j,a}-
\sum_{b\in\mathcal{A}}\phi_{s,j,b}}{|\mathcal{A}|^2}=
\mathbf{U}_{s,j,a}A+\mathbf{V}_{s,j,a}B+F_{s,j,a}$ for a group of given $s\in\mathcal{S}$, $j\in\mathcal{J}$, and $a\in\mathcal{A}$, where every element of $F_{s,j,a}$ is proportional to $e_{w}(s,j,a)$. In this case, Eq.~(\ref{eqb3}) can be written by
\begin{equation}\label{eqb5}
   \begin{split}
    \pi(s,j,a;\theta,\beta)&=1/2+\frac{|\mathcal{A}|\theta^T\phi_{s,j,a}-
    \sum_{b\in\mathcal{A}}\theta^T\phi_{s,j,b}}{|\mathcal{A}|^2}
    \beta+o(\beta),\\
    &=1/2+\left[\underbrace{\theta^T\mathbf{U}_{s,j,a}A+
\theta^T\mathbf{V}_{s,j,a}B}_{m(s,j,a;\theta)}
+\underbrace{\theta^TF_{s,j,a}}_{\mathrm{c}(s,j,a;\theta)}\right]\beta+o(\beta), \\
&=1/2+[m(s,j,a;\theta)+\mathrm{c}(s,j,a;\theta)]\beta+o(\beta).
  \end{split}
\end{equation}
In particular, if one defines a function $\bar{p}_{u,v}(\pi(s,j,a;\theta^*,\beta))$ by
\begin{equation}\label{eqb6}
\begin{split}
\bar{p}_{u,v}(\pi(s,j,a;\theta^*,\beta))=\sum_{j\in\mathcal{J}}
  \left\{
  \begin{array}{ll}
    p_Cp_{C,j}\pi(s,j,C;\theta^*,\beta)+p_Dp_{D, j}
\pi(s,j,D;\theta^*,\beta), & \hbox{for } v=u; \\
    p_Cp_{C, j}\pi(s,j,D;\theta^*,\beta), & \hbox{for } v=u-1;\\
    p_Dp_{D, j}\pi(s,j,C;\theta^*,\beta), & \hbox{for } v=u+1; \\
    0, & \hbox{otherwise};
  \end{array}
\right.
\end{split}
\end{equation}
then $p^{*}_{u,v}$ can be given by $p^{*}_{u,v}=\sum_{s\in\mathcal{S}}d^{\pi}(s)
\bar{p}_{u,v}(\pi(s,j,a;\theta^*,\beta))$ based on Eq.~(\ref{eq9}). Substituting Eq.~(\ref{eqb5}) into it eventually leads to
\begin{equation}\label{eqb7}
p^{*}_{u,v}=\sum_{s\in\mathcal{S}}d^{\pi}(s)
\bar{p}_{u,v}
(1/2+[m(s,j,a;\theta^*)+\mathrm{c}(s,j,a;\theta^*)]\beta+o(\beta)).
\end{equation}
\par
To obtain the long-run distribution $X=[x_n]_{1\times (N+1)}, n\in\mathcal{N}$, we start to figure out the equations $X(P^*-I)=\mathbf{0}_{N+1}$ and $\sum_{n\in\mathcal{N}}x_n=1$. First, we note that $P^*$ is a stochastic matrix because $\sum_{v\in \mathcal{N}}p_{u,v}^*=\sum_{s\in\mathcal{S}}d^{\pi}(s)\sum_{v\in \mathcal{N}}\bar{p}_{u,v}(\pi(s,j,a;\theta^*,\beta))=
\sum_{s\in\mathcal{S}}d^{\pi}(s)=1$ for $\forall u \in\mathcal{N}$ and $p^{*}_{u,v}\geq 0$ for $\forall u,v \in\mathcal{N}$. Moreover, $P^*$ is primitive because any two states of the Markov chain described by the probability transition matrix $P^*$ are accessible to each other, i.e., $(P^*)^\kappa>0$ for some positive integers $\kappa$. Then, the Perron-Frobenius theorem~\cite{saloff1997lectures} ensures that $1$ is its largest eigenvalue and the corresponding eigenvector with all entries summing to $1$ is the unique stationary distribution that we want to seek. That is, the solution to equations $X(P^*-I)=\mathbf{0}_{N+1}$ and $\sum_{n\in\mathcal{N}}x_n=1$ will have only one degree of freedom. Without loss of generality, we assume that it is $x_N$. Then, there will exist a set of coefficients, $h_\varsigma$, $\varsigma=0,1,2,\ldots,N-1$, such that
\begin{equation}\label{eqb8}
  x_\varsigma=h_\varsigma x_N,\text{ for } \varsigma=0,1,2,\ldots,N-1.
\end{equation}
Furthermore, since the stationary distribution is a vector with all entries summing to $1$, we have
\begin{equation}\label{eqb9}
  x_N(1+h_0+h_1+\ldots+h_{N-1})=1.
\end{equation}
\par
On the other hand, based on Eqs.~(\ref{eqb6}) and (\ref{eqb7}), we can find that after performing the Gaussian elimination, the elements of the reduced matrix of $P^*-I$ will be polynomials of $\beta$, and whenever we have a degree $\kappa$ term in $\beta$, it must be accompanied by a degree $\kappa$ term in $\sum_{s\in\mathcal{S}}d^{\pi}(s)[m(s,j,a;\theta^*)+\mathrm{c}(s,j,a;\theta^*)]$. In view of the nature of the Gaussian elimination, it implies that all $h_\varsigma, \varsigma=0,1,2,\ldots,N-1$ will be rational functions of $\beta$. Then, based on Eqs.~(\ref{eqb8}) and (\ref{eqb9}), we can obtain that every $x_n$, $n\in\mathcal{N}$, will be a rational function of $\beta$.  Without loss of generality, we write it in an irreducible form by
\begin{equation}
   x_n=\frac{\ell_{0n}+\ell_{1n}\beta+o(\beta)}
  {\lambda_{0n}+\lambda_{1n}\beta+o(\beta)}, \forall n\in\mathcal{N},
\end{equation}
where $\ell_{0n}$ and $\lambda_{0n}$ are constant terms which are independent of $\sum_{s\in\mathcal{S}}d^{\pi}(s)m(s,j,a;\theta^*)$ and $\sum_{s\in\mathcal{S}}d^{\pi}(s)\mathrm{c}(s,j,a;\theta^*)$, whereas $\ell_{1n}$ and $\lambda_{1n}$ are linear combinations of $\sum_{s\in\mathcal{S}}d^{\pi}(s)m(s,j,a;\theta^*)$ and $\sum_{s\in\mathcal{S}}d^{\pi}(s)\mathrm{c}(s,j,a;\theta^*)$, respectively. Accordingly, the first-order derivative of $x_n$ at
$\beta=0$ can be given by
\begin{equation}
  x_n'(0)=\frac{\ell_{1n}\lambda_{0n}
-\ell_{0n}\lambda_{1n}}{\lambda_{0n}^2}, \forall n\in\mathcal{N}.
\end{equation}
From this equation, we know that $x_n'(0)$ will be a linear combination of $\sum_{s\in\mathcal{S}}d^{\pi}(s)m(s,j,a;\theta^*)$ and $\sum_{s\in\mathcal{S}}d^{\pi}(s)\mathrm{c}(s,j,a;\theta^*)$. That is, there will exist a set of coefficients, $k_{j,a}$ and $g_{j,a}$, such that
\begin{equation}\label{eqb12}
  \sum_{n\in \mathcal{N}}nx_n'(0)=\sum_{j\in\mathcal{J}}\sum_{a\in\mathcal{A}}
[k_{j,a}\sum_{s\in\mathcal{S}}d^{\pi}(s)m(s,j,a;\theta^*)
+g_{j,a}\sum_{s\in\mathcal{S}}d^{\pi}(s)\mathrm{c}(s,j,a;\theta^*)].
\end{equation}
Based on Eqs.~(\ref{eqb1}) and (\ref{eqb2}), it follows that $\langle x_C\rangle>1/2$ under weak selection if and only if
\begin{equation}\label{eqb13}
\sum_{j\in\mathcal{J}}\sum_{a\in\mathcal{A}}
[k_{j,a}\sum_{s\in\mathcal{S}}d^{\pi}(s)m(s,j,a;\theta^*)
+g_{j,a}\sum_{s\in\mathcal{S}}d^{\pi}(s)\mathrm{c}(s,j,a;\theta^*)]
>0.
\end{equation}

\subsection{The final condition}
Based on the results we obtain above, we here give the final condition for cooperation to be favored in a simple form of payoff entries, $\mathrm{a}_j(s)$ and $\mathrm{b}_j(s)$. First, based on Eq.~(\ref{eqb5}), we substitute $m(s,j,a;\theta^*)=\theta^{*T}\mathbf{U}_{s,j,a}A+\theta^{*T}\mathbf{V}_{s,j,a}B$ and $\mathrm{c}(s,j,a;\theta^*)=\theta^{*T}F_{s,j,a}$ into Eq.~(\ref{eqb12}). It leads to
\begin{equation}\label{eqb14}
\begin{split}
&\sum_{n\in \mathcal{N}}nx_n'(0)=\sum_{j\in\mathcal{J}}\sum_{a\in\mathcal{A}}
\left\{k_{j,a}\sum_{s\in\mathcal{S}}d^{\pi}(s)\left[
\theta^{*T}\mathbf{U}_{s,j,a}A+\theta^{*T}\mathbf{V}_{s,j,a}B\right]
+g_{j,a}\sum_{s\in\mathcal{S}}d^{\pi}(s)\theta^{*T}F_{s,j,a}\right\}.
\end{split}
\end{equation}
Let $\Phi_s=\sum_{j\in\mathcal{J}}\sum_{a\in\mathcal{A}}k_{j,a}\mathbf{U}_{s,j,a}$, $\Psi_s=\sum_{j\in\mathcal{J}}\sum_{a\in\mathcal{A}}k_{j,a}\mathbf{V}_{s,j,a}$, and $e=\sum_{j\in\mathcal{J}}\sum_{a\in\mathcal{A}}g_{j,a}\sum_{s\in\mathcal{S}}\\
d^{\pi}(s)\theta^{*T}F_{s,j,a}$. Then, Eq.~(\ref{eqb14}) is changed to
\begin{equation}\label{eqb15}
 \sum_{n\in \mathcal{N}}nx_n'(0)=
\sum_{s\in\mathcal{S}}d^{\pi}(s)\theta^{*T}\Phi_s
A+\sum_{s\in\mathcal{S}}d^{\pi}(s)\theta^{*T}\Psi_s
B+e,
\end{equation}
and meanwhile condition~(\ref{eqb13}) is changed to
\begin{equation}\label{eqb16}
 \sum_{s\in\mathcal{S}}d^{\pi}(s)\theta^{*T}\Phi_s
A+\sum_{s\in\mathcal{S}}d^{\pi}(s)\theta^{*T}\Psi_s
B+e>0.
\end{equation}
\par
Particularly, it is noteworthy that the strategic updating process described in our model is symmetric for the two actions, $C$ and $D$. That is, if we relabel the action notations (i.e., exchanging $C$ and $D$) and swap their corresponding payoff entries in Table~\ref{table1} (i.e., exchanging $\mathrm{a}_{d-1-j}(s)$ and $\mathrm{b}_j(s)$ for all $j\in\mathcal{J}$ and $s\in\mathcal{S}$), it will result in symmetric dynamics~\cite{wu2013dynamic,tarnita2009strategy}. The reason is that the unique difference between action $C$ and $D$ is fully captured by the payoff table, and the population structure as well as the reinforcement learning algorithm do not introduce any distinctions between these two actions. Then, based on Eqs.~(\ref{eqb1}), (\ref{eqb2}), and (\ref{eqb15}), we know that the average abundance of $D$ players, $\langle x_D\rangle$, in the population, after enforcing the above swapping operations, can be given by
\begin{equation}\label{eqb17}
\begin{split}
  \langle x_D\rangle&=\frac{1}{2}+\frac{\beta}{N}
\left[\sum_{s\in\mathcal{S}}d^{\pi}(s)\theta^{*T}\Phi_s
B+\sum_{s\in\mathcal{S}}d^{\pi}(s)\theta^{*T}\Psi_s
A+e\right]+o(\beta).
\end{split}
\end{equation}
Accordingly, the average abundance of $C$ players in the population is
\begin{equation}\label{eqb18}
\begin{split}
 \langle x_C\rangle&=1- \langle x_D\rangle \\
&=\frac{1}{2}-\frac{\beta}{N}
\left[\sum_{s\in\mathcal{S}}d^{\pi}(s)\theta^{*T}\Phi_s
B+\sum_{s\in\mathcal{S}}d^{\pi}(s)\theta^{*T}\Psi_s
A+e\right]+o(\beta).
\end{split}
\end{equation}
It follows that $\langle x_C\rangle>1/2$ if and only if
\begin{equation}\label{eqb19}
-\sum_{s\in\mathcal{S}}d^{\pi}(s)\theta^{*T}\Phi_s
B-\sum_{s\in\mathcal{S}}d^{\pi}(s)\theta^{*T}\Psi_s
A-e>0.
\end{equation}
Since both inequalities (\ref{eqb16}) and (\ref{eqb19}) are the condition under which $\langle x_C\rangle$ is greater than $1/2$ and they hold for any stationary distribution $d^{\pi}(s)$, $\theta^{*T}\in\mathbb{R}^L$, $A\in\mathbb{R}^{dM}$, and $B\in\mathbb{R}^{dM}$, there must exist a positive scale factor $\zeta>0$ such that $\Phi_s=-\zeta\Psi_s$,
$\Psi_s=-\zeta\Phi_s$,
and $e=-\zeta e$, for any $s\in\mathcal{S}$. Then, we have $\zeta=1$ and $e=0$. As a result, $\langle x_C\rangle>1/2$ if and only if
\begin{equation}\label{eqb20}
  \begin{split}
&\sum_{s\in\mathcal{S}}d^{\pi}(s)\theta^{*T}\Phi_s
A+\sum_{s\in\mathcal{S}}d^{\pi}(s)\theta^{*T}\Psi_s
B+e\\
&=\sum_{s\in\mathcal{S}}d^{\pi}(s)\theta^{*T}\Phi_s
A-\sum_{s\in\mathcal{S}}d^{\pi}(s)\theta^{*T}\Phi_s
B\\
&=\sum_{s\in\mathcal{S}}d^{\pi}(s)\theta^{*T}\Phi_s
(A-B) \\
&>0.
\end{split}
\end{equation}

\section{Finite well-mixed populations and structured populations}
In this part, using the mean-field approximation, we derive the condition for cooperation to be favored in two specific population structures, finite well-mixed populations and structured populations, in the limit of weak selection and large population size.
\subsection{Finite well-mixed populations}
In a finite and well-mixed population, the interactive links of individuals are described by a complete graph. To obtain the stationary proportion of $C$ players in the population, we first calculate the probabilities that the number of $C$ players at time $\tau$, $n_{\tau}$, increases by one and decreases by one, which are given by
\begin{equation}
  T^+(n_{\tau}=n)=\sum_{s\in\mathcal{S}}
\overbrace{Pr\{s_{\tau}=s|s_0,\pi\}}^{\substack{\text{the environmental}\\ \text{state is $s$ at time $\tau$}}}
\overbrace{\frac{N-n}{N}}^{\substack{\text{a focal player}\\ \text{ of $D$ is chosen}}}
\sum_{j=0}^{d-1}
\overbrace{\frac{{n\choose j}{N-1-n\choose d-1-j}}{{N-1\choose d-1}}}^{\substack{\text{there are $j$ co-}\\ \text{players of $C$}}}
\overbrace{\pi(s,j,C;\theta_{\tau},\beta)}^{\substack{\text{the focal player}\\ \text{ changes action to $C$}}},
\end{equation}
and
\begin{equation}
  T^-(n_{\tau}=n)=\sum_{s\in\mathcal{S}}
\overbrace{Pr\{s_{\tau}=s|s_0,\pi\}}^{\substack{\text{the environmental}\\ \text{state is $s$ at time $\tau$}}}
\overbrace{\frac{n}{N}}^{\substack{\text{a focal player}\\ \text{ of $C$ is chosen}}}
\sum_{j=0}^{d-1}
\overbrace{\frac{{n-1\choose j}{N-n\choose d-1-j}}{{N-1\choose d-1}}}^{\substack{\text{there are $j$ co-}\\ \text{players of $C$}}}
\overbrace{\pi(s,j,D;\theta_{\tau},\beta)}^{\substack{\text{the focal player}\\ \text{ changes action to $D$}}},
\end{equation}
respectively.
Then, the master equation describing the evolutionary dynamics of the number of $C$ players can be given by
\begin{equation}\label{eqc3}
\begin{split}
& P(n_{\tau+1}=n)-  P(n_{\tau}=n) =T^+(n_{\tau}=n-1)P(n_{\tau}=n-1)
\\
  &+T^-(n_{\tau}=n+1)P(n_{\tau}=n+1)
-[T^+(n_{\tau}=n)+T^-(n_{\tau}=n)]
P(n_{\tau}=n),
\end{split}
\end{equation}
where $P(n_{\tau}=n)$ is the probability that the population contains $n$ players of $C$ at time $\tau$. Next, to perform the diffusion approximation~\cite{van1992stochastic,traulsen2005coevolutionary} to the master equation for sufficiently large population size $N\gg 1$, we scale $\tau$ by $N$, denoted by $t=\tau/N$, and introduce $y=n/N$ and the probability density $f(y,t)=NP(n_{\tau}=n)$. Then, the master equation (\ref{eqc3}) is changed to
\begin{equation}\label{eqsi33}
  \begin{split}
& f(y,t+1/N)-  f(y,t) =T^+(y-1/N)f(y-1/N,t)
\\
  &+T^-(y+1/N)f(y+1/N,t)
-[T^+(y)+T^-(y)]
f(y,t).
\end{split}
\end{equation}
For $N\gg 1$, we expand the probability densities and transition probabilities in Eq.~(\ref{eqsi33}) in a Taylor series at $y$ and $t$. Neglecting higher order terms in $1/N$ , we obtain
\begin{equation}
  \frac{\partial}{\partial t}f(y,t)=-\frac{\partial}{\partial y}[\varphi(y)f(y,t)]
+\frac{1}{2}\frac{\partial^2}{\partial y^2}[\psi^2(y)f(y,t)],
\end{equation}
where $\varphi(y)=T^+(y)-T^-(y)$ is the drift term and $\psi(y)=\sqrt{[T^+(y)+T^-(y)]/N}$ is the diffusion term. Note that this equation has the form of the Fokker-Planck equation~\cite{van1992stochastic}, and the internal noise of the system is not correlated because successive update steps are mutually independent. Then, the
It\^{o} calculus can be applied to derive the Langevin equation~\cite{van1992stochastic} $\dot{y}=\varphi(y)+\psi(y)\varrho$, where $\varrho$ is the uncorrelated Gaussian noise. In particular, it is worth noting that for
sufficiently large population size $N\rightarrow\infty$, the diffusion term $\psi(y)$ will vanish. Therefore, in this case, the dynamics describing the evolution of the proportion of $C$ players reduce to a deterministic differential equation,
\begin{equation}\label{eq33}
\begin{split}
   \dot{y}&=T^+(y)-T^-(y) \\
&=\sum_{s\in\mathcal{S}}Pr\{s_t=s|s_0,\pi\}
\left[(1-y)\sum_{j=0}^{d-1}H(y,j)   \pi(s,j,C;\theta_t,\beta)\right.\\
&\left.-y\sum_{j=0}^{d-1}H(y,j)
\pi(s,j,D;\theta_t,\beta)
\right],
\end{split}
\end{equation}
where $H(y,j)\triangleq{d-1\choose j}y^{j}(1-y)^{d-1-j}$ is the binomial distribution used for approximating the hypergeometric distribution,
\begin{equation}
  H(y,j)\approx
\frac{{n\choose j}{N-1-n\choose d-1-j}}{{N-1\choose d-1}}
\approx
\frac{{n-1\choose j}{N-n\choose d-1-j}}{{N-1\choose d-1}},
\end{equation}
for sufficiently large $N$.
\par
On the other hand, note that when $\theta_t$ is updated via the actor-critic reinforcement learning algorithm, Eq.~(\ref{eq7}), it will almost surely converge to the equilibrium of the following dynamic equations by applying stochastic approximation theory~\cite{borkar1997stochastic},
\begin{equation}\label{eqc8}
\begin{split}
  \varepsilon \dot{w}_t&=-\frac{\partial \parallel Q^{\pi}(s,j,a)-f_{w_t}(s,j,a)\parallel_{\pi}^2}{\partial w_t}, \\
\dot{\theta}_t&=\frac{\partial \rho(\pi)}{\partial \theta_t},
\end{split}
\end{equation}
where $\varepsilon$ is a sufficiently small perturbation parameter. Combining these two evolutionary processes together, a complete expression of the system dynamics for $N\gg 1$ can be given by
\begin{equation}\label{eq}
\begin{split}
  \left\{
  \begin{array}{ll}
   \dot{y}
&=\sum_{s\in\mathcal{S}}Pr\{s_t=s|s_0,\pi\}
\sum_{j=0}^{d-1}H(y,j)\\
&\left[(1-y)\pi(s,j,C;\theta_t,\beta)-y
\pi(s,j,D;\theta_t,\beta)\right], \\
  \varepsilon \dot{w}_t&=-\frac{\partial \parallel Q^{\pi}(s,j,a)-f_{w_t}(s,j,a)\parallel_{\pi}^2}{\partial w_t}, \\
\dot{\theta}_t&=\frac{\partial \rho(\pi)}{\partial \theta_t}.
  \end{array}
\right.
\end{split}
\end{equation}
From this equation, one can find that the dynamics of the reinforcement learning algorithm are independent of the evolution of the proportion of $C$ players, but it is not true in turn. Based on the analysis in the Supporting Information SI.1, we know that $\theta_t$ will converge to a local optimum of $\rho(\pi)$, denoted by $\theta^*$ (i.e., the solution to $\dot{\theta}_t=\frac{\partial \rho(\pi)}{\partial \theta_t}=0$). Then, we can obtain the equilibrium of the proportion of $C$ players, $y^*$, by solving $\dot{y}=0$ when $\theta_t$ has converged and the environment has evolved to the stationary state, that is,
\begin{equation}\label{eqc10}
  \sum_{s\in\mathcal{S}}d^{\pi}(s)
\sum_{j=0}^{d-1}H(y^*,j)\left[(1-y^*)   \pi(s,j,C;\theta^*,\beta)-y^*
\pi(s,j,D;\theta^*,\beta)
\right]=0,
\end{equation}
where $d^\pi(s)=\lim_{t\rightarrow \infty}Pr\{s_t=s|s_0,\pi\}$ is applied. Because the average proportion of $C$ players in the population always keeps one half under neutral selection (i.e., $\beta=0$), the equilibrium of the proportion of $C$ players, $y^*$, can be written by $1/2$ plus some perturbations in the limit of weak selection, $y^*=1/2+\xi\beta+o(\beta)$, where $\xi\triangleq\frac{\partial y^*}{\partial \beta}|_{\beta=0}$. Substituting $y^*=1/2+\xi\beta+o(\beta)$ and $\pi(s,j,a;\theta^*,\beta)=1/2+\frac{2\theta^{*T}\phi_{s,j,a}-
\sum_{b\in\mathcal{A}}\theta^{*T}\phi_{s,j,b}}{4}
\beta+o(\beta)$, for $\forall a\in\mathcal{A}$ into Eq.~(\ref{eqc10}), we then get
\begin{equation}
  \sum_{s\in\mathcal{S}}d^{\pi}(s)
\sum_{j=0}^{d-1}H(y^*,j)
\left[ -\xi \beta +\frac{\theta^{*T}}{4}(\phi_{s,j,C}
-\phi_{s,j,D})
\beta+o(\beta)\right]=0.
\end{equation}
Solving $\xi$ leads to
\begin{equation}
\begin{split}
\xi&= \sum_{s\in\mathcal{S}}d^{\pi}(s)
\sum_{j=0}^{d-1}H(y^*,j)
\frac{\theta^{*T}}{4}\left[\phi_{s,j,C}-\phi_{s,j,D}
\right], \\
&=\sum_{s\in\mathcal{S}}d^{\pi}(s)
\sum_{j=0}^{d-1}\left[{d-1\choose j}\frac{1}{2^{d-1}}+O(\beta) \right]
\frac{\theta^{*T}}{4}\left[\phi_{s,j,C}-\phi_{s,j,D}
\right], \\
&=\sum_{s\in\mathcal{S}}d^{\pi}(s)
\sum_{j=0}^{d-1}{d-1\choose j}\frac{1}{2^{d+1}}
\theta^{*T}\left[\phi_{s,j,C}-\phi_{s,j,D}
\right].
\end{split}
\end{equation}
It follows that under weak selection and for large population size, $y^*=1/2+\xi\beta+o(\beta)>1/2$ if and only if
\begin{equation}\label{eqc13}
\sum_{s\in\mathcal{S}}d^{\pi}(s)
\sum_{j=0}^{d-1}{d-1\choose j}\frac{1}{2^{d+1}}
\theta^{*T}\left[\phi_{s,j,C}-\phi_{s,j,D}
\right]>0.
\end{equation}
\subsection{Structured populations}
Here, we proceed to consider a structured population where the interactive links of individuals are described by a regular graph with node degree $d-1$. To capture the evolutionary dynamics of the system, we begin with defining the number of $C$ players by $n_t$, and the proportion of $C$ players in the population by $P_C(t)=n_t/N$, at time $t$. Then, the probabilities that the proportion of $C$ players increases and decreases by $1/N$ when $n_t=n$, can be given by
\begin{equation}
  Pr(\Delta P_C(t)=\frac{1}{N})=\sum_{s\in\mathcal{S}}
\overbrace{Pr\{s_{t}=s|s_0,\pi\}}^{\substack{\text{the environmental}\\ \text{state is $s$ at time $t$}}}
\overbrace{\frac{N-n}{N}}^{\substack{\text{a focal player}\\ \text{ of $D$ is chosen}}}
\sum_{j=0}^{d-1}
\overbrace{G(q_{C|D},q_{D|D},j)}^{\substack{\text{there are $j$ co-}\\ \text{players of $C$}}}
\overbrace{\pi(s,j,C;\theta_t,\beta)}^{\substack{\text{the focal player}\\ \text{ changes action to $C$}}},
\end{equation}
and
\begin{equation}
  Pr(\Delta P_C(t)=-\frac{1}{N})=\sum_{s\in\mathcal{S}}
\overbrace{Pr\{s_{t}=s|s_0,\pi\}}^{\substack{\text{the environmental}\\ \text{state is $s$ at time $t$}}}
\overbrace{\frac{n}{N}}^{\substack{\text{a focal player}\\ \text{ of $C$ is chosen}}}
\sum_{j=0}^{d-1}
\overbrace{G(q_{C|C},q_{D|C},j)}^{\substack{\text{there are $j$ co-}\\ \text{players of $C$}}}
\overbrace{\pi(s,j,D;\theta_t,\beta)}^{\substack{\text{the focal player}\\ \text{ changes action to $D$}}},
\end{equation}
respectively, where $G(q_{C|a},q_{D|a},j)\triangleq {d-1\choose j}q_{C|a}^{j}q_{D|a}^{d-1-j}$, $\forall a\in\mathcal{A}$, and $q_{Z|Y}$ denotes the conditional probability for a $Y$ player to have a $Z$ neighbor on average, $\forall Y,Z\in \mathcal{A}$, and satisfies $q_{C|a}+q_{D|a}=1$ for $\forall a\in\mathcal{A}$.
Since only the focal player can revise its action in one unit of time, the time derivative of $P_C(t)$ can be given by~\cite{ohtsuki2006simple}
\begin{equation}\label{eqc16}
\begin{split}
\dot{P}_{C}(t)&=\frac{1}{N}Pr(\Delta P_{C}(t)=\frac{1}{N})-\frac{1}{N}Pr(\Delta P_{C}(t)=-\frac{1}{N}),\\
&=\frac{1}{N}\sum_{s\in\mathcal{S}}
Pr\{s_{t}=s|s_0,\pi\}\left[(1-P_C(t))\sum_{j=0}^{d-1}
G(q_{C|D},q_{D|D},j)\pi(s,j,C;\theta_t,\beta) \right.\\
&\left.-P_C(t)\sum_{j=0}^{d-1}
G(q_{C|C},q_{D|C},j)\pi(s,j,D;\theta_t,\beta)
\right].
\end{split}
\end{equation}
Combining with the dynamics of the reinforcement learning, Eqs.~(\ref{eqc8}), we can then obtain the system dynamics given by
\begin{equation}\label{eq}
\begin{split}
  \left\{
  \begin{array}{ll}
\dot{P}_{C}(t)&=\frac{1}{N}\sum_{s\in\mathcal{S}}
Pr\{s_{t}=s|s_0,\pi\}\left[(1-P_C(t))\sum_{j=0}^{d-1}
G(q_{C|D},q_{D|D},j)\pi(s,j,C;\theta_t,\beta) \right.\\
&\left.-P_C(t)\sum_{j=0}^{d-1}
G(q_{C|C},q_{D|C},j)\pi(s,j,D;\theta_t,\beta)
\right], \\
\varepsilon \dot{w}_t&=-\frac{\partial \parallel Q^{\pi}(s,j,a)-f_{w_t}(s,j,a)\parallel_{\pi}^2}{\partial w_t}, \\
\dot{\theta}_t&=\frac{\partial \rho(\pi)}{\partial \theta_t}.
  \end{array}
\right.
\end{split}
\end{equation}
Similar to the situation in well-mixed populations, the learning dynamics are independent of the evolution of the proportion of $C$ players. Thus, we can obtain the equilibrium of the proportion of $C$ players, $P_C^*$, in structured populations, by solving $\dot{P}_{C}(t)=0$ when $\theta_t$ has converged and the environment has evolved to the stationary state, that is,
\begin{equation}\label{eqc18}
\begin{split}
\sum_{s\in\mathcal{S}}
d^{\pi}(s)\left[(1-P_C^*)\sum_{j=0}^{d-1}
G(q_{C|D},q_{D|D},j)\pi(s,j,C;\theta^*,\beta) \right.\\
\left.-P_C^*\sum_{j=0}^{d-1}
G(q_{C|C},q_{D|C},j)\pi(s,j,D;\theta^*,\beta)
\right]=0,
\end{split}
\end{equation}
where $d^\pi(s)=\lim_{t\rightarrow \infty}Pr\{s_t=s|s_0,\pi\}$ is applied. Under weak selection, we first expand $P_C^*$ and $\pi(s,j,a;\theta^*,\beta)$ in the first-order Taylor series, $P_C^*=1/2+\epsilon \beta+o(\beta)$ and $\pi(s,j,a;\theta^*,\beta)=1/2+\frac{2\theta^{*T}\phi_{s,j,a}-
    \sum_{b\in\mathcal{A}}\theta^{*T}\phi_{s,j,b}}{4}
    \beta+o(\beta), \forall a\in\mathcal{A}$, where  $\epsilon\triangleq\frac{\partial P_C^*}{\partial \beta}|_{\beta=0}$.
Substituting them into Eq.~(\ref{eqc18}) and solving $\epsilon$, we then get
\begin{equation}
\epsilon= \sum_{s\in\mathcal{S}}d^{\pi}(s)
\sum_{j=0}^{d-1}{d-1\choose j}\frac{1}{2^{d+1}}
\theta^{*T}\left[\phi_{s,j,C}-\phi_{s,j,D}
\right].
\end{equation}
It follows that, under weak selection and for large population size, $P_C^*=1/2+\epsilon \beta+o(\beta)>1/2$ if and only if
\begin{equation}\label{eqc20}
\sum_{s\in\mathcal{S}}d^{\pi}(s)
\sum_{j=0}^{d-1}{d-1\choose j}\frac{1}{2^{d+1}}
\theta^{*T}\left[\phi_{s,j,C}-\phi_{s,j,D}
\right]>0.
\end{equation}
Interestingly, this condition is identical to the inequality (\ref{eqc13}), which, once again, confirms the previous finding that there are no differences in the final condition for cooperation to be favored under weak selection between complete graphs and regular graphs~\cite{wu2018individualised}.
\par
Particularly, we here highlight that conditions (\ref{eqc13}) and (\ref{eqc20}) are the specific forms of condition (\ref{eqb20}) in finite well-mixed and structured populations, respectively. First, note that $\phi_{s,j,a}$ is a $L$-dimension vector which is handcrafted to characterize the feature when the focal player takes action $a$ given the environmental state $s$ and the number of $C$ players $j$ among its $d-1$ co-players, and $L$ is usually chosen to be much smaller than the dimension of the environmental state $M$ for the sake of reducing dimensions. Then, for any a choice of $\phi_{s,j,a}$, we can always find two sets of matrices, $\Phi_{s,j}\in\mathbb{R}^{L\times dM}$ and $\Psi_{s,j}\in\mathbb{R}^{L\times dM}$, such that $\phi_{s,j,C}=\Phi_{s,j}(A+\hbar\mathbf{1})$ and $-\phi_{s,j,D}=\Psi_{s,j}(B+\hbar\mathbf{1})$, where $\mathbf{1}$ is the $dM$-dimension vector with all element $1$, and $\hbar$ is a large constant chosen to ensure that every element of $A+\hbar\mathbf{1}$ and $B+\hbar\mathbf{1}$ is greater than zero. Let $\Phi_s=\sum_{j=0}^{d-1}{d-1\choose j}\frac{1}{2^{d+1}}\Phi_{s,j}$ and $\Psi_s=\sum_{j=0}^{d-1}{d-1\choose j}\frac{1}{2^{d+1}}\Psi_{s,j}$. Then, both Eq.~(\ref{eqc13}) and Eq.~(\ref{eqc20}) are changed to $\sum_{s\in\mathcal{S}}d^{\pi}(s)
\theta^{*T}\left[\Phi_s(A+\hbar\mathbf{1})
+\Psi_s(B+\hbar\mathbf{1})\right]>0$. Using the symmetric property of the strategic updating and via a similar computational process as Eqs.~(\ref{eqb17}) -- (\ref{eqb19}), we can obtain $\Phi_s=-\Psi_s$. As a result, the proportion of $C$ players in the population is higher than that of $D$ players if and only if $\sum_{s\in\mathcal{S}}d^{\pi}(s)
\theta^{*T}\Phi_s(A-B)>0$.

\section{Smoothed best response updating and aspiration-based updating}
In this section, we apply our model framework to study two non-learning updating processes, the smoothed best response of the Fermi form and the aspiration-based rule, and derive the condition for cooperation to be favored in the limit of weak selection.
\par
If the focal player in our model updates actions per step via the smoothed best response of the Fermi form, then the probability for the focal player to choose action $a\in\mathcal{A}$ is given by Eq.~(\ref{eq12}). Under weak selection, its first-order Taylor expansion is given by
\begin{equation}\label{eqd1}
  \pi(s,j,a;\beta)=1/2+\frac{\mathcal{R}_{s,j}^a-\mathcal{R}_{s,j}^b}{4}\beta+o(\beta),\ \forall s\in \mathcal{S}, j\in\mathcal{J}, a,b(\neq a)\in \mathcal{A}.
\end{equation}
\par
In contrast, if the aspiration-based updating rule is adopted, the probability for the focal player to switch to the new action $a\in\mathcal{A}$ will be given by Eq.~(\ref{eq13}). Analogously, in the limit of weak selection, we can give its first-order Taylor expansion by
\begin{equation}\label{eqd2}
  \pi(s,j,a;\beta)=1/2+\frac{\mathcal{R}_{s,j}^a-\mathcal{E}}{4}\beta+o(\beta),\ \forall s\in \mathcal{S}, j\in\mathcal{J}, a\in \mathcal{A}.
\end{equation}
In particular, if the aspiration level is not a fixed constant but a time-varying value, we will have an adapting aspiration scheme,
\begin{equation}\label{eqd3}
  \pi(s,j,a;\beta)=\frac{1}
  {1+e^{-\beta [\mathcal{R}_{s,j}^a-\mathcal{E}(t)]}},\ \forall s\in \mathcal{S}, j\in\mathcal{J}, a\in \mathcal{A}.
\end{equation}
To update the aspiration level, a simple rule $\Delta\mathcal{E}(t)=\omega[\Omega-\mathcal{E}(t)]$ can be adopted, where $0<\omega<1$ denotes the updating step-size and $\Omega$ specifies a desired aspiration. Compared with the constant aspiration, this rule means that there are some noises or fluctuations for the desired aspiration $\Omega$ during the evolution of cooperation, as the aspiration level will asymptotically converge to the unique stable equilibrium, $\lim_{t\rightarrow +\infty}\mathcal{E}(t)=\Omega$. In particular, if $\Omega$ is set to be $r_{t+1}$, it will recover the learning rule used in adapting aspiration dynamics~\cite{posch1999efficiency}. Once again, in the equilibrium state, we give its first-order Taylor expansion under weak selection by
\begin{equation}\label{eqd4}
    \pi(s,j,a;\beta)=1/2+\frac{\mathcal{R}_{s,j}^a-\Omega}{4}\beta+o(\beta),\ \forall s\in \mathcal{S}, j\in\mathcal{J}, a\in \mathcal{A}.
\end{equation}
\par
Since $\mathcal{R}_{s,j}^a=\mathrm{a}_j(s)$ if $a=C$ and $\mathcal{R}_{s,j}^a=\mathrm{b}_j(s)$ if $a=D$, as shown in Eqs.~(\ref{eq4}), we can always find a set of coefficients $\tilde{\mu}(a)$ and $\tilde{\nu}(a)$ for Eq.~(\ref{eqd1}), (\ref{eqd2}), or (\ref{eqd4}) such that
\begin{equation}
    \pi(s,j,a;\beta)=1/2+\left[\underbrace{\tilde{\mu}(a)\mathrm{a}_{j}(s)
+\tilde{\nu}(a)\mathrm{b}_{j}(s)}_{\tilde{m}(s,j,a)}+\tilde{\mathrm{c}}(s,j,a)\right]\beta+o(\beta),
\end{equation}
for any given $s\in \mathcal{S}$, $j\in\mathcal{J}$, and $a\in \mathcal{A}$, where $\tilde{\mathrm{c}}(s,j,a)$ is a constant. Then, the transition probability $p^{*}_{u,v}$ can be given by $p^{*}_{u,v}=\sum_{s\in\mathcal{S}}d^{\pi}(s)
\bar{p}_{u,v}
(1/2+[\tilde{m}(s,j,a)+\tilde{\mathrm{c}}(s,j,a)]\beta+o(\beta))$, based on Eqs.~(\ref{eqb5}) and (\ref{eqb7}). According to Eq.~(\ref{eqb12}), it follows that there exists a set of coefficients $\tilde{k}_{j,a}$ and $\tilde{g}_{j,a}$ such that
\begin{equation}\label{eq}
\begin{split}
\sum_{n\in \mathcal{N}}nx_n'(0)&=\sum_{j\in\mathcal{J}}\sum_{a\in\mathcal{A}}
[\tilde{k}_{j,a}\sum_{s\in\mathcal{S}}d^{\pi}(s)\tilde{m}(s,j,a)
+\tilde{g}_{j,a}\sum_{s\in\mathcal{S}}d^{\pi}(s)\tilde{\mathrm{c}}(s,j,a)] \\
&=\sum_{j\in\mathcal{J}}\sum_{a\in\mathcal{A}}
\tilde{k}_{j,a}\sum_{s\in\mathcal{S}}d^{\pi}(s)
[\tilde{\mu}(a)\mathrm{a}_{j}(s)+\tilde{\nu}(a)\mathrm{b}_{j}(s)] \\
&+\underbrace{\sum_{j\in\mathcal{J}}\sum_{a\in\mathcal{A}}\tilde{g}_{j,a}
\sum_{s\in\mathcal{S}}d^{\pi}(s)\tilde{\mathrm{c}}(s,j,a)}_{\tilde{\mathbf{c}}} \\
&=
\sum_{s\in\mathcal{S}}d^{\pi}(s)\sum_{j\in\mathcal{J}}\left[
\underbrace{\sum_{a\in\mathcal{A}}\tilde{k}_{j,a}\tilde{\mu}(a)}_{\sigma_{j}}\mathrm{a}_{j}(s) +\underbrace{\sum_{a\in\mathcal{A}}\tilde{k}_{j,a}\tilde{\nu}(a)}_{\delta_{j}}\mathrm{b}_{j}(s)
\right]+\tilde{\mathbf{c}}  \\
&=\sum_{s\in\mathcal{S}}d^{\pi}(s)\sum_{j\in\mathcal{J}}\left[
\sigma_j\mathrm{a}_j(s)+\delta_j\mathrm{b}_j(s)
\right]+\tilde{\mathbf{c}}.
\end{split}
\end{equation}
Then, based on Eqs.~(\ref{eqb1}) and (\ref{eqb2}), we have $\langle x_C\rangle>1/2$ under weak selection if and only if
\begin{equation}\label{eqd8}
  \sum_{s\in\mathcal{S}}d^{\pi}(s)\sum_{j\in\mathcal{J}}\left[
\sigma_j\mathrm{a}_j(s)+\delta_j\mathrm{b}_j(s)
\right]+\tilde{\mathbf{c}}>0.
\end{equation}
Finally, in view of the symmetric property of these two update rules and via a similar computational process as Eqs.~(\ref{eqb17}) -- (\ref{eqb19}), we can get $\sigma_j=-\delta_{d-1-j}$ and $\tilde{\mathbf{c}}=0$. As a result, inequality~(\ref{eqd8}) is changed to
\begin{equation}
  \sum_{s\in\mathcal{S}}d^{\pi}(s)\sum_{j\in\mathcal{J}}
\sigma_j[\mathrm{a}_j(s)-\mathrm{b}_{d-1-j}(s)]>0,
\end{equation}
where $\sigma_j$, $\forall j\in\mathcal{J}$, are some coefficients needed to be calculated for the given population structure, and independent of both $\mathrm{a}_j(s)$ and $\mathrm{b}_j(s)$. In particular, for finite well-mixed and structured populations, these coefficients can be obtained by solving Eq.~(\ref{eqc10}) and Eq.~(\ref{eqc18}), respectively. Substituting Eq.~(\ref{eqd1}), (\ref{eqd2}), or (\ref{eqd4}) into them, these $\sigma$-coefficients are given by $\sigma_j={d-1\choose j}/2^{d+1}$ for the smoothed best response of the Fermi form, and $\sigma_j={d-1\choose j}/2^{d+2}$ for the aspiration-based updating rule, in either finite well-mixed populations or structured populations.

\begin{figure}
  \centering
  \includegraphics[width=\hsize]{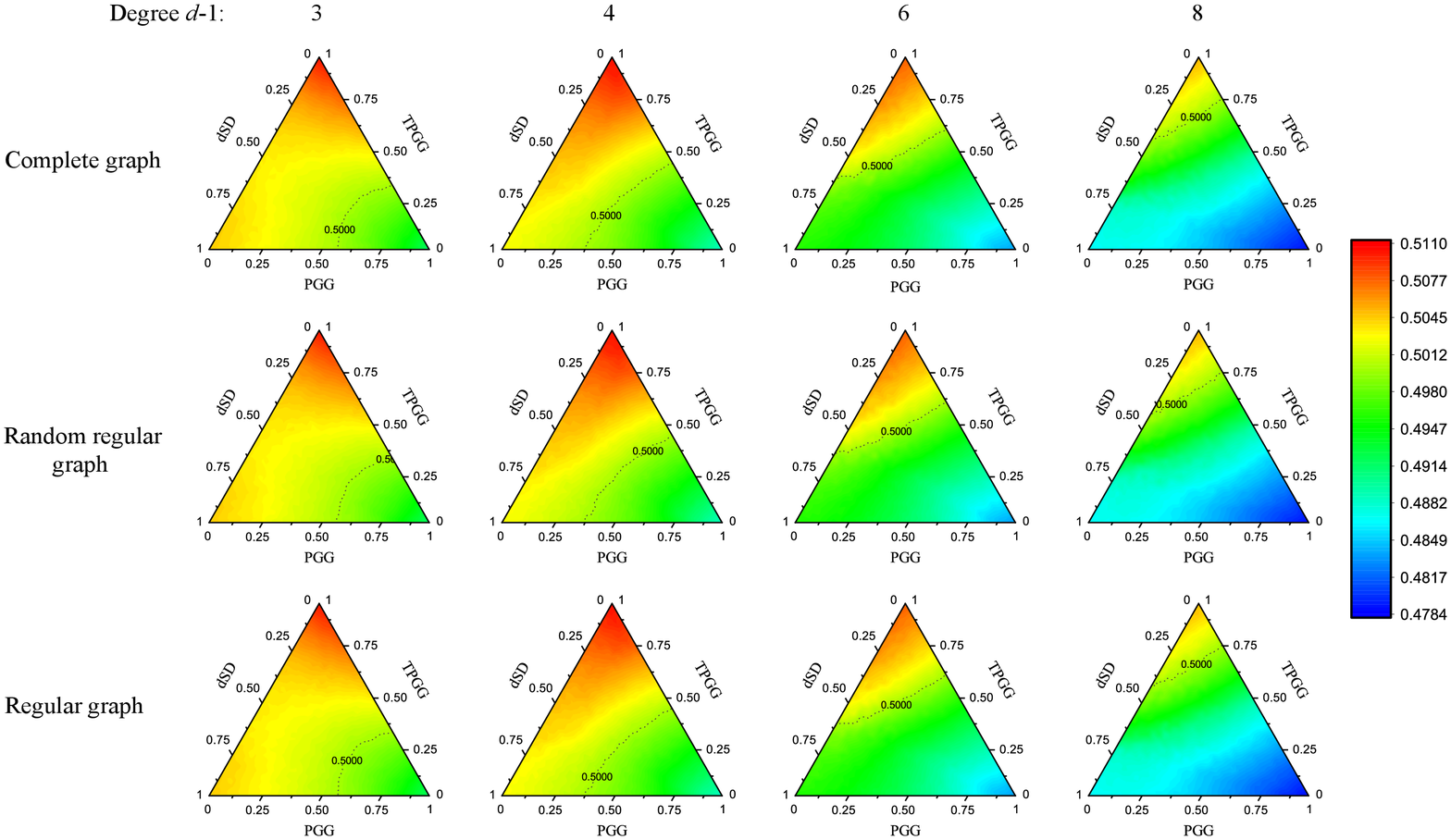}\\
  \caption{Average abundance of $C$ players for different population structures and different network degrees when players
  update actions via the reinforcement learning. Parameter values are the same as in Fig.~\ref{fig3}.}\label{figS1}
\end{figure}

\begin{figure}
  \centering
  \includegraphics[width=\hsize]{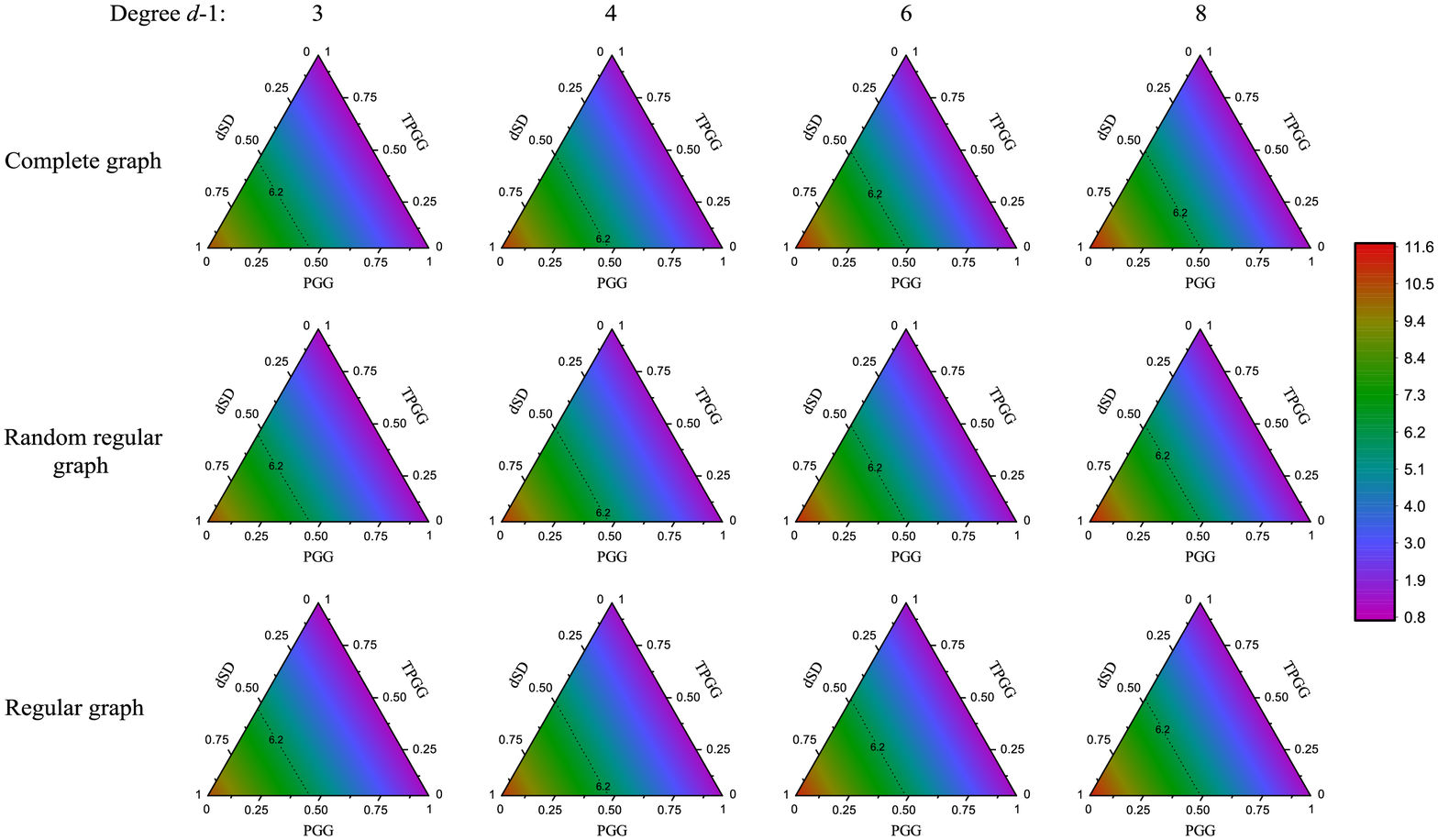}\\
  \caption{Expected payoff per
round for different population structures and different network degrees when players update actions via the reinforcement learning. Parameter values are the same as in Fig.~\ref{fig3}.}\label{figS2}
\end{figure}

\begin{figure}
  \centering
  \includegraphics[width=\hsize]{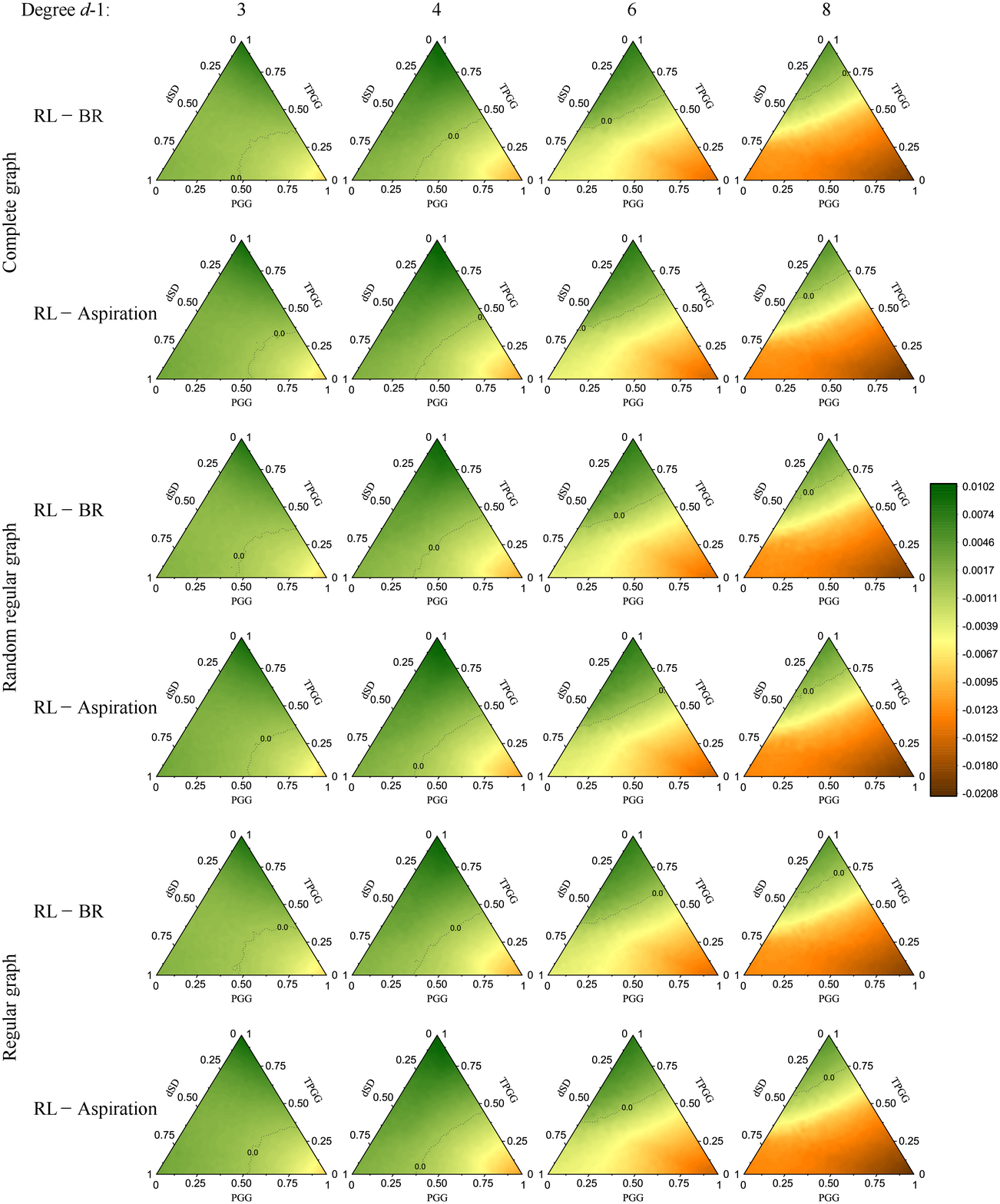}\\
  \caption{Differences in the average abundance of $C$ players between the reinforcement learning (RL) and the non-learning updates -- the smoothed best response (BR) and the aspiration-based rule (Aspiration), for different population structures and different network degrees. Parameter values are the same as in Fig.~\ref{fig3}.}\label{figS3}
\end{figure}

\begin{figure}
  \centering
  \includegraphics[width=\hsize]{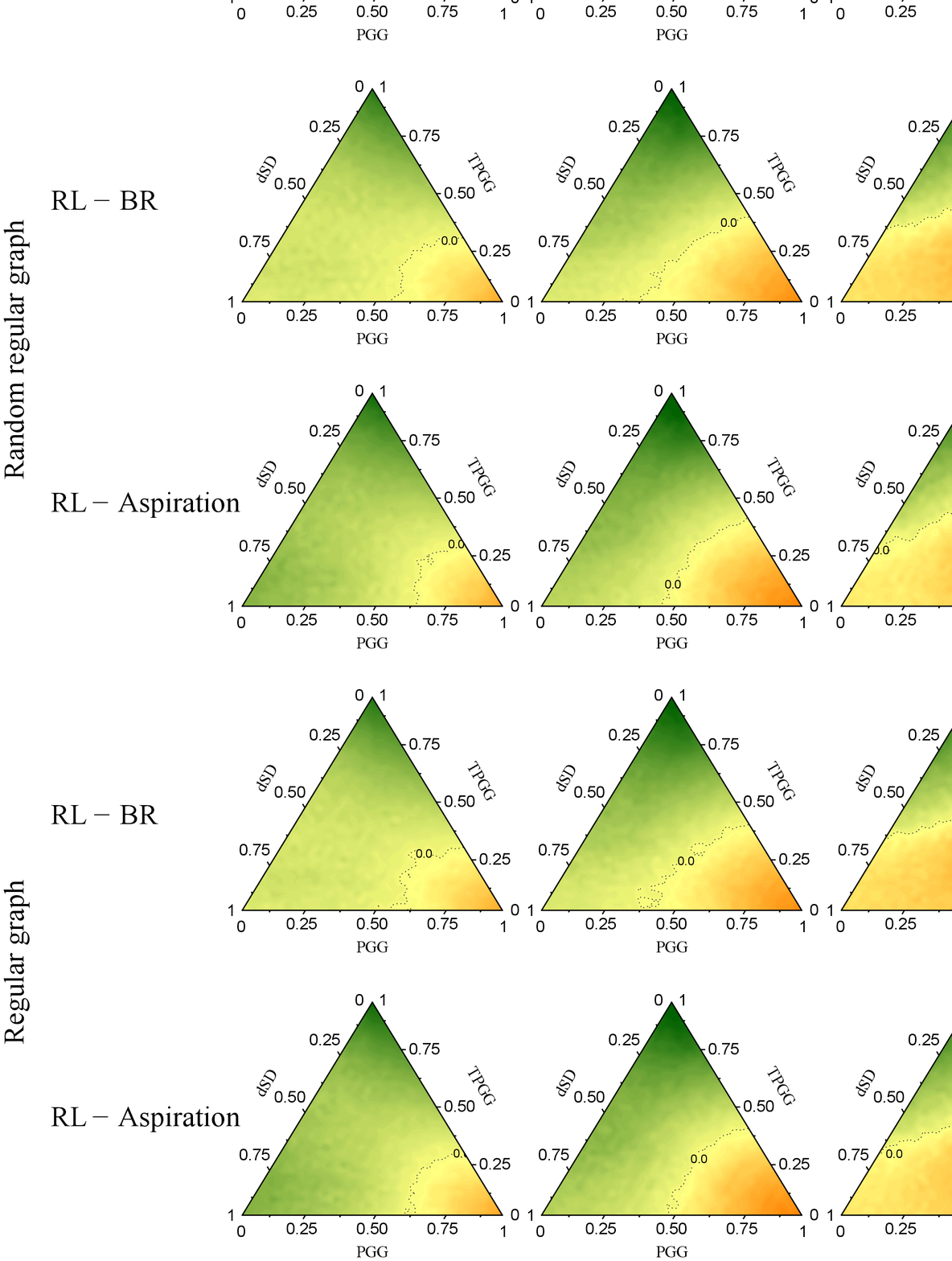}\\
  \caption{Differences in the expected payoff per
round between the reinforcement learning (RL) and the non-learning updates -- the smoothed best response (BR) and the aspiration-based rule (Aspiration), for different population structures and different network degrees. Parameter values are the same as in Fig.~\ref{fig3}.}\label{figS4}
\end{figure}

\begin{figure}
  \centering
  \includegraphics[width=\hsize]{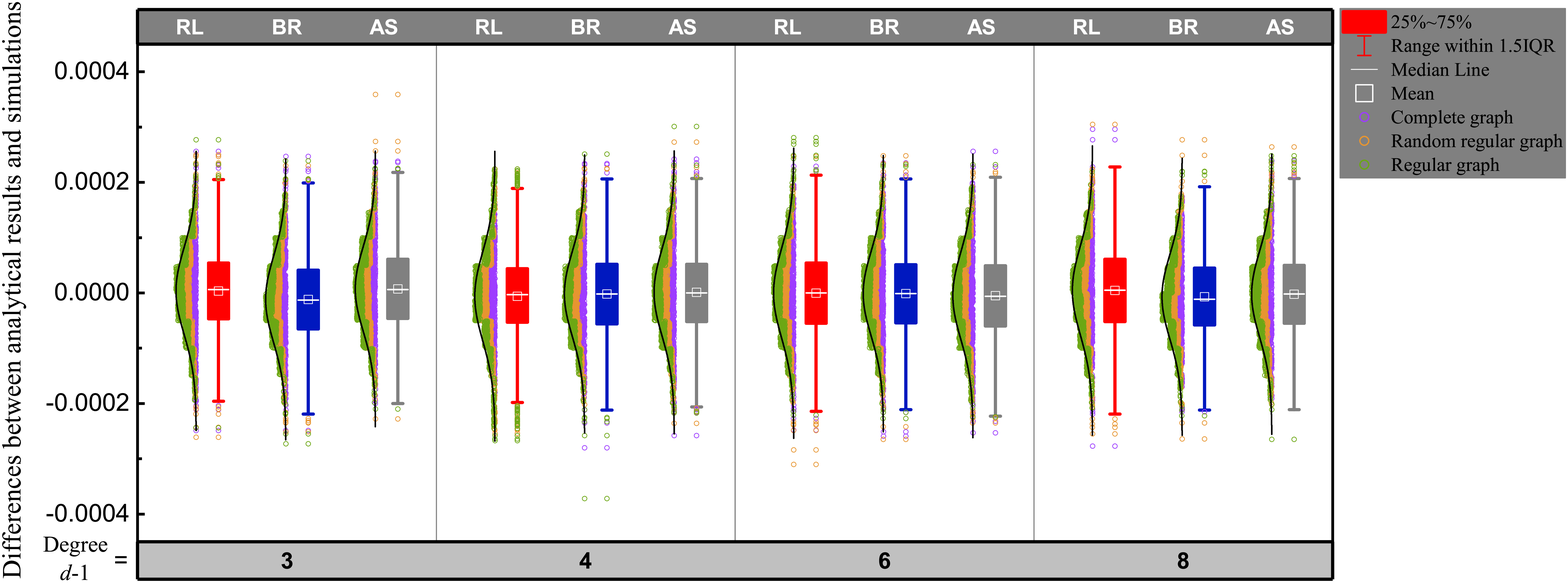}\\
  \caption{Differences between analytical results and simulations for different population structures and different network degrees. RL, BR, and AS represent the reinforcement learning, smoothed best response, and aspiration-based rule, respectively. Circles are the error values between analytical results and simulations obtained in the whole simplex parameter space consisting of the distribution of the PGG, TPGG, and dSD. Parameter values are the same as in Fig.~\ref{fig3}.}\label{figS5}
\end{figure}

\clearpage
\bibliographystyle{elsarticle-num}
\bibliography{apssamp}

\providecommand{\noopsort}[1]{}\providecommand{\singleletter}[1]{#1}%
\begin{thebibliography}{10}
\expandafter\ifx\csname url\endcsname\relax
  \def\url#1{\texttt{#1}}\fi
\expandafter\ifx\csname urlprefix\endcsname\relax\def\urlprefix{URL }\fi
\expandafter\ifx\csname href\endcsname\relax
  \def\href#1#2{#2} \def\path#1{#1}\fi

\bibitem{west2007evolutionary}
S.~A. West, A.~S. Griffin, A.~Gardner, Evolutionary explanations for
  cooperation, Curr. Biol. 17~(16) (2007) R661--R672.

\bibitem{pachauri2010climate}
R.~K. Pachauri, Climate ethics: Essential readings, Oxford University Press,
  Oxford, UK, 2010.

\bibitem{milinski2008collective}
M.~Milinski, R.~D. Sommerfeld, H.-J. Krambeck, F.~A. Reed, J.~Marotzke, The
  collective-risk social dilemma and the prevention of simulated dangerous
  climate change, Proc. Natl. Acad. Sci. USA 105~(7) (2008) 2291--2294.

\bibitem{ostrom2015governing}
E.~Ostrom, Governing the commons: The evolution of institutions for collective
  action, Cambridge University Press, Cambridge, UK, 1990.

\bibitem{colman2006puzzle}
A.~M. Colman, The puzzle of cooperation, Nature 440~(7085) (2006) 744--745.

\bibitem{nowak2006five}
M.~A. Nowak, Five rules for the evolution of cooperation, Science 314~(5805)
  (2006) 1560--1563.

\bibitem{dawkins2016selfish}
R.~Dawkins, The selfish gene, Oxford University Press, Oxford, UK, 2016.

\bibitem{smith1982evolution}
J.~M. Smith, Evolution and the theory of games, Cambridge University Press,
  Cambridge, UK, 1982.

\bibitem{hofbauer1998evolutionary}
J.~Hofbauer, K.~Sigmund, Evolutionary games and population dynamics, Cambridge
  University Press, Cambridge, UK, 1998.

\bibitem{hamilton1964genetical}
W.~D. Hamilton, The genetical evolution of social behaviour
  \uppercase\expandafter{\romannumeral1} and
  \uppercase\expandafter{\romannumeral2}, J. Theor. Biol. 7~(1) (1964) 1--52.

\bibitem{szabo2007evolutionary}
G.~Szab{\'o}, G.~Fath, Evolutionary games on graphs, Phys. Rep. 446~(4-6)
  (2007) 97--216.

\bibitem{archetti2012game}
M.~Archetti, I.~Scheuring, Game theory of public goods in one-shot social
  dilemmas without assortment, J. Theor. Biol. 299 (2012) 9--20.

\bibitem{gokhale2010evolutionary}
C.~S. Gokhale, A.~Traulsen, Evolutionary games in the multiverse, Proc. Natl.
  Acad. Sci. USA 107~(12) (2010) 5500--5504.

\bibitem{tarnita2011multiple}
C.~E. Tarnita, N.~Wage, M.~A. Nowak, Multiple strategies in structured
  populations, Proc. Natl. Acad. Sci. USA 108~(6) (2011) 2334--2337.

\bibitem{wu2013dynamic}
B.~Wu, A.~Traulsen, C.~S. Gokhale, Dynamic properties of evolutionary
  multi-player games in finite populations, Games 4~(2) (2013) 182--199.

\bibitem{pena2016ordering}
J.~Pe{\~n}a, B.~Wu, A.~Traulsen, Ordering structured populations in multiplayer
  cooperation games, J. R. Soc. Interface 13~(114) (2016) 20150881.

\bibitem{mcavoy2016structure}
A.~McAvoy, C.~Hauert, Structure coefficients and strategy selection in
  multiplayer games, J. Math. Biol. 72~(1-2) (2016) 203--238.

\bibitem{hardin1968tragedy}
G.~Hardin, The tragedy of the commons, Science 162~(3859) (1968) 1243--1248.

\bibitem{weitz2016oscillating}
J.~S. Weitz, C.~Eksin, K.~Paarporn, S.~P. Brown, W.~C. Ratcliff, An oscillating
  tragedy of the commons in replicator dynamics with game-environment feedback,
  Proc. Natl. Acad. Sci. USA 113~(47) (2016) E7518--E7525.

\bibitem{hilbe2018evolution}
C.~Hilbe, {\v{S}}.~{\v{S}}imsa, K.~Chatterjee, M.~A. Nowak, Evolution of
  cooperation in stochastic games, Nature 559~(7713) (2018) 246--249.

\bibitem{estrela2019environmentally}
S.~Estrela, E.~Libby, J.~Van~Cleve, F.~D{\'e}barre, M.~Deforet, W.~R. Harcombe,
  J.~Pe{\~n}a, S.~P. Brown, M.~E. Hochberg, Environmentally mediated social
  dilemmas, Trends Ecol. Evol. 34~(1) (2019) 6--18.

\bibitem{tilman2020evolutionary}
A.~R. Tilman, J.~B. Plotkin, E.~Ak{\c{c}}ay, Evolutionary games with
  environmental feedbacks, Nat. Commun. 11~(1) (2020) 1--11.

\bibitem{macarthur1970species}
R.~MacArthur, Species packing and competitive equilibrium for many species,
  Theor. Popul. Biol. 1~(1) (1970) 1--11.

\bibitem{levins1968evolution}
R.~Levins, Evolution in changing environments: Some theoretical explorations,
  Princeton University Press, Princeton, New Jersey, USA, 1968.

\bibitem{ROSENBERG20201}
N.~A. Rosenberg, Fifty years of theoretical population biology, Theor. Popul.
  Biol. 133 (2020) 1 -- 12.

\bibitem{ashcroft2014fixation}
P.~Ashcroft, P.~M. Altrock, T.~Galla, Fixation in finite populations evolving
  in fluctuating environments, J. R. Soc. Interface 11~(100) (2014) 20140663.

\bibitem{chen2018punishment}
X.~Chen, A.~Szolnoki, Punishment and inspection for governing the commons in a
  feedback-evolving game, PLoS Comput. Biol. 14~(7) (2018) e1006347.

\bibitem{Su19Evolutionary}
Q.~Su, A.~McAvoy, L.~Wang, M.~A. Nowak, Evolutionary dynamics with game
  transitions, Proc. Natl. Acad. Sci. USA 116~(51) (2019) 25398--25404.

\bibitem{hauert2019asymmetric}
C.~Hauert, C.~Saade, A.~McAvoy, Asymmetric evolutionary games with
  environmental feedback, J. Theor. Biol. 462 (2019) 347--360.

\bibitem{hashimoto2006unpredictability}
K.~Hashimoto, Unpredictability induced by unfocused games in evolutionary game
  dynamics, J. Theor. Biol. 241~(3) (2006) 669--675.

\bibitem{venkateswaran2019evolutionary}
V.~R. Venkateswaran, C.~S. Gokhale, Evolutionary dynamics of complex multiple
  games, Proc. R. Soc. B 286~(1905) (2019) 20190900.

\bibitem{stewart2014collapse}
A.~J. Stewart, J.~B. Plotkin, Collapse of cooperation in evolving games, Proc.
  Natl. Acad. Sci. USA 111~(49) (2014) 17558--17563.

\bibitem{akiyama2000dynamical}
E.~Akiyama, K.~Kaneko, Dynamical systems game theory and dynamics of games,
  Physica D 147~(3-4) (2000) 221--258.

\bibitem{shapley1953stochastic}
L.~S. Shapley, Stochastic games, Proc. Natl. Acad. Sci. USA 39~(10) (1953)
  1095--1100.

\bibitem{neyman2003stochastic}
A.~Neyman, S.~Sorin (Eds.), Stochastic games and applications, Kluwer Academic
  Press, Dordrecht, The Netherlands, 2003.

\bibitem{meyers2002fighting}
L.~A. Meyers, J.~J. Bull, Fighting change with change: Adaptive variation in an
  uncertain world, Trends Ecol. Evol. 17~(12) (2002) 551--557.

\bibitem{ballare1990far}
C.~L. Ballar{\'e}, A.~L. Scopel, R.~A. S{\'a}nchez, Far-red radiation reflected
  from adjacent leaves: An early signal of competition in plant canopies,
  Science 247~(4940) (1990) 329--332.

\bibitem{danforth1999emergence}
B.~N. Danforth, Emergence dynamics and bet hedging in a desert bee, perdita
  portalis, Proc. R. Soc. B 266~(1432) (1999) 1985--1994.

\bibitem{thorndike1970animal}
E.~L. Thorndike, Animal Intelligence: Experimental studies, Macmillan, New
  York, USA, 1911.

\bibitem{dayan2008reinforcement}
Y.~Niv, Reinforcement learning in the brain, J. Math. Psychol. 53~(3) (2009)
  139--154.

\bibitem{sutton2018reinforcement}
R.~S. Sutton, A.~G. Barto, Reinforcement learning: An introduction, MIT Press,
  Cambridge, Massachusetts, USA, 2018.

\bibitem{bu2008comprehensive}
L.~Busoniu, R.~Babuska, B.~De~Schutter, A comprehensive survey of multiagent
  reinforcement learning, IEEE Trans. Syst. Man Cybernet. C 38~(2) (2008)
  156--172.

\bibitem{fudenberg1998theory}
D.~Fudenberg, D.~Levine, The theory of learning in games, MIT Press, Cambridge,
  Massachusetts, USA, 1998.

\bibitem{camerer2011behavioral}
C.~F. Camerer, Behavioral game theory: Experiments in strategic interaction,
  Princeton University Press, Princeton, New Jersey, USA, 2011.

\bibitem{nowak2004emergence}
M.~A. Nowak, A.~Sasaki, C.~Taylor, D.~Fudenberg, Emergence of cooperation and
  evolutionary stability in finite populations, Nature 428~(6983) (2004)
  646--650.

\bibitem{sato2005stability}
Y.~Sato, E.~Akiyama, J.~P. Crutchfield, Stability and diversity in collective
  adaptation, Physica D 210~(1-2) (2005) 21--57.

\bibitem{sutton2000policy}
R.~S. Sutton, D.~A. McAllester, S.~P. Singh, Y.~Mansour, Policy gradient
  methods for reinforcement learning with function approximation, in: Adv.
  Neural Inf. Process. Syst., Vol.~12, 1999, pp. 1057--1063.

\bibitem{konda2000actor}
V.~R. Konda, J.~N. Tsitsiklis, Actor-critic algorithms, in: Adv. Neural Inf.
  Process. Syst., Vol.~12, 1999, pp. 1008--1014.

\bibitem{borkar1997stochastic}
V.~S. Borkar, Stochastic approximation with two time scales, Systems Control
  Lett. 29~(5) (1997) 291--294.

\bibitem{bertsekas1996neuro}
D.~P. Bertsekas, J.~N. Tsitsiklis, Neuro-dynamic programming, Athena
  Scientific, Belmont, Massachusetts, USA, 1996.

\bibitem{isaacson1976markov}
D.~L. Isaacson, R.~W. Madsen, Markov chains theory and applications, John Wiley
  \& Sons, New York, USA, 1976.

\bibitem{bowerman1974nonstationary}
B.~L. Bowerman, Nonstationary markov decision processes and related topics in
  nonstationary markov chains, Ph.D. thesis, Iowa State University (1974).

\bibitem{ibsen2015computational}
R.~Ibsen-Jensen, K.~Chatterjee, M.~A. Nowak, Computational complexity of
  ecological and evolutionary spatial dynamics, Proc. Natl. Acad. Sci. USA
  112~(51) (2015) 15636--15641.

\bibitem{tuyls2003selection}
K.~Tuyls, K.~Verbeeck, T.~Lenaerts, A selection-mutation model for q-learning
  in multi-agent systems, in: Proc. of 2nd Intl. Conf. on Autonomous Agents and
  Multiagent Systems (AAMAS 2003), ACM, 2003, pp. 693--700.

\bibitem{tarnita2009strategy}
C.~E. Tarnita, H.~Ohtsuki, T.~Antal, F.~Fu, M.~A. Nowak, Strategy selection in
  structured populations, J. Theor. Biol. 259~(3) (2009) 570--581.

\bibitem{barfuss2020caring}
W.~Barfuss, J.~F. Donges, V.~V. Vasconcelos, J.~Kurths, S.~A. Levin, Caring for
  the future can turn tragedy into comedy for long-term collective action under
  risk of collapse, Proc. Natl. Acad. Sci. USA 117~(23) (2020) 12915--12922.

\bibitem{du2014aspiration}
J.~Du, B.~Wu, P.~M. Altrock, L.~Wang, Aspiration dynamics of multi-player games
  in finite populations, J. R. Soc. Interface 11~(94) (2014) 20140077.

\bibitem{souza2009evolution}
M.~O. Souza, J.~M. Pacheco, F.~C. Santos, Evolution of cooperation under
  n-person snowdrift games, J. Theor. Biol. 260~(4) (2009) 581--588.

\bibitem{wu2018individualised}
B.~Wu, L.~Zhou, Individualised aspiration dynamics: Calculation by proofs, PLoS
  Comput. Biol. 14~(9) (2018) e1006035.

\bibitem{pacheco2009evolutionary}
J.~M. Pacheco, F.~C. Santos, M.~O. Souza, B.~Skyrms, Evolutionary dynamics of
  collective action in n-person stag hunt dilemmas, Proc. R. Soc. B 276~(1655)
  (2009) 315--321.

\bibitem{fehr2003nature}
E.~Fehr, U.~Fischbacher, The nature of human altruism, Nature 425~(6960) (2003)
  785--791.

\bibitem{sigmund2010social}
K.~Sigmund, H.~De~Silva, A.~Traulsen, C.~Hauert, Social learning promotes
  institutions for governing the commons, Nature 466~(7308) (2010) 861--863.

\bibitem{perc2017statistical}
M.~Perc, J.~J. Jordan, D.~G. Rand, Z.~Wang, S.~Boccaletti, A.~Szolnoki,
  Statistical physics of human cooperation, Phys. Rep. 687 (2017) 1--51.

\bibitem{crandall2018cooperating}
J.~W. Crandall, M.~Oudah, F.~Ishowo-Oloko, et~al., Cooperating with machines,
  Nat. Commun. 9~(1) (2018) 1--12.

\bibitem{rahwan2019machine}
I.~Rahwan, M.~Cebrian, N.~Obradovich, et~al., Machine behaviour, Nature
  568~(7753) (2019) 477--486.

\bibitem{macy2002learning}
M.~W. Macy, A.~Flache, Learning dynamics in social dilemmas, Proc. Natl. Acad.
  Sci. USA 99~(suppl 3) (2002) 7229--7236.

\bibitem{sato2002chaos}
Y.~Sato, E.~Akiyama, J.~D. Farmer, Chaos in learning a simple two-person game,
  Proc. Natl. Acad. Sci. USA 99~(7) (2002) 4748--4751.

\bibitem{galla2013complex}
T.~Galla, J.~D. Farmer, Complex dynamics in learning complicated games, Proc.
  Natl. Acad. Sci. USA 110~(4) (2013) 1232--1236.

\bibitem{barfuss2019deterministic}
W.~Barfuss, J.~F. Donges, J.~Kurths, Deterministic limit of temporal difference
  reinforcement learning for stochastic games, Phys. Rev. E 99~(4) (2019)
  043305.

\bibitem{bloembergen2015evolutionary}
D.~Bloembergen, K.~Tuyls, D.~Hennes, M.~Kaisers, Evolutionary dynamics of
  multi-agent learning: A survey, J. Artif. Intell. Res. 53 (2015) 659--697.

\bibitem{dridi2014learning}
S.~Dridi, L.~Lehmann, On learning dynamics underlying the evolution of learning
  rules, Theor. Popul. Biol. 91 (2014) 20--36.

\bibitem{dridi2018learning}
S.~Dridi, E.~Ak{\c{c}}ay, Learning to cooperate: The evolution of social
  rewards in repeated interactions, Am. Nat. 191~(1) (2018) 58--73.

\bibitem{khalvati2019modeling}
K.~Khalvati, S.~A. Park, S.~Mirbagheri, R.~Philippe, M.~Sestito, J.-C. Dreher,
  R.~P. Rao, Modeling other minds: Bayesian inference explains human choices in
  group decision-making, Sci. Adv. 5~(11) (2019) eaax8783.

\bibitem{ramazi2016networks}
P.~Ramazi, J.~Riehl, M.~Cao, Networks of conforming or nonconforming
  individuals tend to reach satisfactory decisions, Proc. Natl. Acad. Sci. USA
  113~(46) (2016) 12985--12990.

\bibitem{kaelbling1998planning}
L.~P. Kaelbling, M.~L. Littman, A.~R. Cassandra, Planning and acting in
  partially observable stochastic domains, Artif. Intell. 101~(1-2) (1998)
  99--134.

\bibitem{saloff1997lectures}
L.~Saloff-Coste, Lectures on finite markov chains, in: Lectures on probability
  theory and statistics, Springer, 1997, pp. 301--413.

\bibitem{van1992stochastic}
N.~G. Van~Kampen, Stochastic processes in physics and chemistry, 2nd Edition,
  Elsevier, Amsterdam, The Netherlands, 1997.

\bibitem{traulsen2005coevolutionary}
A.~Traulsen, J.~C. Claussen, C.~Hauert, Coevolutionary dynamics: from finite to
  infinite populations, Phys. Rev. Lett. 95~(23) (2005) 238701.

\bibitem{ohtsuki2006simple}
H.~Ohtsuki, C.~Hauert, E.~Lieberman, M.~A. Nowak, A simple rule for the
  evolution of cooperation on graphs and social networks, Nature 441~(7092)
  (2006) 502--505.

\bibitem{posch1999efficiency}
M.~Posch, A.~Pichler, K.~Sigmund, The efficiency of adapting aspiration levels,
  Proc. R. Soc. Lond. B 266~(1427) (1999) 1427--1435.

\end{thebibliography}

\end{document}